\documentclass[reprint,letterpaper,footinbib,showpacs,amsmath,amssymb,aps,pre]{revtex4-1}
\usepackage{graphicx}
\usepackage{bm}
\usepackage{natbib}

\usepackage{CJK}

\begin{document}
\begin{CJK*}{UTF8}{gbsn}
\title{Chemical sensing by cell-surface chemoreceptor arrays: the roles of receptor cooperativity and adaptation}
\author{Jin Yang (杨劲)}
\email{jinyang2004@gmail.com}
\affiliation{Chinese Academy of Sciences -- Max Plank Society Partner Institute and Key Laboratory for Computational Biology,\\ Shanghai Institutes for Biological Sciences, Shanghai 200031, China}

\begin{abstract}
Most sensory cells use cross-membrane chemoreceptors to detect chemical signals in the environment. The biochemical properties and spatial organization of chemoreceptors play important roles in achieving and maintaining sensitivity and accuracy of chemical sensing. Here we investigate the effects of receptor cooperativity and adaptation on the limits of gradient sensing. We study a single cell with aggregated chemoreceptor arrays on the cell surface and derive general formula to the limits for gradient sensing from the uncertainty of instantaneous receptor activity. In comparison to independent receptors, we find that cooperativity by non-adaptative receptors could significantly lower the sensing limit in a chemical concentration range determined by the biochemical properties of ligand-receptor binding and ligand-induced receptor activity. Cooperativity by adaptative receptors are beneficial to gradient sensing within a broad range of background concentrations. Our results also show that isotropic receptor aggregate layout on the cell surface represents an optimal configuration to gradient sensing.
\end{abstract}

\maketitle
\end{CJK*}

\section{Introduction}

Cellular sensory systems can detect temporal and spatial changes of environmental signals. For example, bacteria sense chemical gradient by motion that translates spatial chemical concentration asymmetry into temporal asymmetry. Larger chemotactic eukaryotes are able to sense the spatial gradients of chemoattractants across the cell dimension. A primary task for a sensory cell is to respond to small chemical concentration changes or shallow chemical gradients with sufficient accuracy under stochastic noises. Berg and Purcell, in their classic study~\cite{berg1977physics}, showed that the fundamental physical limit of concentration sensing is set by the uncertainty of ligand diffusion to a sensory cell regardless of biochemical details in a sensing mechanism employed by the cell. This result was generalized by Bialek and Setayeshgar~\cite{bialek2005physical} using the fluctuation-dissipation theorem, and was also applicable to gradient sensing~\cite{endres2008accuracy,*endres2009accuracy}.

Ligand detection by membrane-bound chemoreceptors is the first step for chemical sensing in a vast majority of sensory systems of living cells. Uncertainty of ligand-receptor binding and stochastic dynamics of downstream cellular signaling may introduce additional noises. Reducing such noise may help a system to operate near the fundamental Berg-Purcell limit. However, the exact sensing limit under proper ligand-receptor interaction remains elusive. In particular, the effect of receptor cooperativity is controversial~\cite{bialek2008cooperativity,hu2010physical,aquino2011optimal,skoge2011dynamics}. Receptor cooperativity can sensitize the receptor response to small signal changes, but in the meanwhile it also amplifies stochastic fluctuations in signal. Using the Monod-Wyman-Changeux (MWC) model, Bialek and Setayeshgar~\cite{bialek2008cooperativity} showed that receptor cooperativity helps to lower the threshold of concentration sensing to approach the Berg-Purcell limit set by ligand diffusion. Hu {\it et al.}~\cite{hu2010physical}, by an Ising-type model, showed that receptor cooperativity improves gradient sensing within a shortened dynamic range of background concentrations. Most recently, Skoge {\it et al.}~\cite{skoge2011dynamics} studied chemoreceptor activity by a dynamic Ising model and showed that receptor cooperativity could slow down receptor activity and thus may not improve the signal-to-noise ratio due to reduced response time.

Another crucial mechanism found in many cellular sensory systems is adaptation, which maintains the sensitivity of a system to varying levels of environmental signals. It is known that adaptation tunes kinetics of chemoreceptors at the molecular level. For the example of the better understood {\it E. coli}, the activity of receptors are modulated by multisite receptor methylation and demethylation. However, the effect of receptor cooperativity to chemical sensing limit under the context of receptor adaptation is yet to be examined. 

\begin{figure}[b]
\centering
\includegraphics[scale=1]{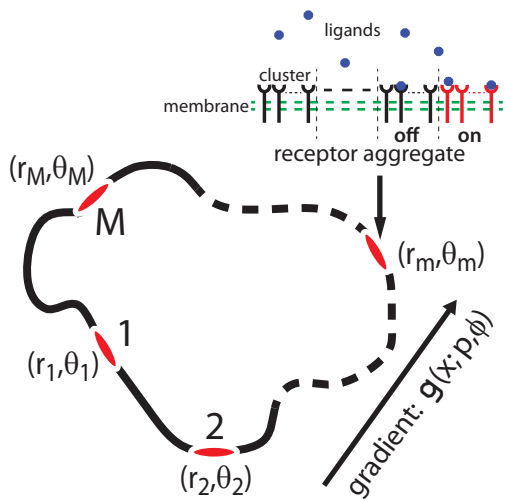}
\caption{\label{fig:scheme} (color online). A 2-D chemotactic cell with $M$ chemoreceptor aggregates (patches labeled with polar coordinates $(r_m,\theta_m)$) distributed on the cell surface (enclosed curve) under a gradient field $\mathcal{G}(\mathbf{x}; p,\phi)$ indicated by the arrow. Each receptor aggregate has $N_m, m=1,...,M$, cooperative receptor clusters of identical size $n$.}
\end{figure}

Using a theoretical model, here we study the role of receptor cooperativity on the accuracy of chemical sensing under the influence of receptor adaptation. We consider a sensory cell with a hierarchical organization of cell-surface chemoreceptors. Receptor cooperativity is described by the classic MWC model~\cite{monod1965nature,*changeux2012}, which was originally developed to explain allosteric regulation of multi-subunit proteins and has been widely used to model chemoreceptor coupling in bacteria~\cite{sourjik2004functional,mello2005allosteric,keymer2006chemosensing,skoge2006receptor}. We derive formula for the limit of gradient sensing based on the uncertainty of instantaneous receptor states at the equilibrium. Our results show that cooperativity by non-adaptative receptors reduces instantaneous noise of receptor states within a limited range of background concentrations determined by the biochemical parameters of receptor dynamics and ligand-receptor binding. In contrast, cooperativity by adaptative receptors improves the sensing accuracy across a wide dynamic range. We also show that the layout of receptor aggregates on the cell surface significantly affects the sensing limit with the isotropic layout being the optimal. Although the sensing limit is sensitive to the cell orientation under anisotropic aggregate layouts, the effect of receptor cooperativity is invariant.

\section{Theory}
\subsection{A two-dimensional cell model}
We consider a model (Fig.~\ref{fig:scheme}) for a two-dimensional chemotactic cell subject to a chemoattractant gradient field $\mathcal{G}(\mathbf{x}; p,\phi)$, where $\mathbf{x}$ is the spatial coordinate, and $p$ and $\phi$ are the steepness and direction of the gradient. The steepness is defined in a polar coordinate system as: $p\equiv\frac{r_0}{c_0}\frac{dc}{dr}$, which is a normalized concentration change across a reference distance $r_0$ along the direction of the gradient. The cell has $M$ receptor aggregates distributed at distinct locations on the cell surface. Cooperative receptors form clusters of size $n$ and each aggregate contains a number of such independent and non-interacting receptor clusters. Assuming aggregate $m$ has $N_m$ receptor clusters, we have the total number of receptor monomers in a cell as: $N_{\rm tot}=n\sum_{m=1}^MN_m$. This hierarchical organization of receptors on the cell membrane is general and can be parameterized to study a cell that has aggregated receptor arrays with ($n>1$) or without cooperativity ($n=1$) or has non-aggregated receptor monomers ($N_m=1$ and $n=1$).

In the absence of ligand, a receptor switches between active (``on") and inactive (``off") state with a free energy difference $\Delta E=E_{\rm off}-E_{\rm on}$ (in units of $k_BT$, where $k_B$ is the Boltzmann constant and $T$ is the absolute temperature). As in an MWC model, coupled receptors in a cluster switch in an all-or-none fashion between the ``on" and ``off" states. A ligand binds to a receptor of the ``on" or ``off" state with a dissociation constant $K_{\rm on}$ or $K_{\rm off}$, respectively. Ligand binding will shift the free energy difference between the two receptor states and such change can be sensed by a cell to measure the ligand concentration. At the equilibrium, a receptor cluster $m$ at the ``on"  or ``off" state bound to $r$ ligands  has the free energy of  $nE_{\rm on}-\ln\left(\frac{[L]_m}{K_{\rm on}}\right)^r$ or $nE_{\rm off}-\ln\left(\frac{[L]_m}{K_{\rm off}}\right)^r$, $r=0,...,n$, respectively, with a multiplicity of $n \choose r$. From Boltzmann distribution, one can obtain the equilibrium probability for a receptor cluster in aggregate $m$ to be active is given by~\cite{mello2005allosteric,keymer2006chemosensing}:
\begin{equation}\label{eq:pm}
P_m = \left[1+\left(e^{-\Delta E}\displaystyle\frac{1+c_m}{1+\alpha c_m}\right)^n\right]^{-1}, \ \ \, m=1,...,M
\end{equation}
where $c_m\equiv [L]_m/K_{\rm off}$ is a normalized ligand concentration. The ligand concentration $[L]_m$ at the location of aggregate $m$ is determined by the gradient field $\mathcal{G}(\mathbf{x}; p,\phi)$. The non-dimensional coefficient $\alpha\equiv K_{\rm off}/K_{\rm on}$.

At the equilibrium, an instantaneous configuration of receptor cluster states measures the chemical concentrations around the cell. Such configuration fluctuates in time due to randomness of ligand-receptor binding and stochasticity in receptor state switching, which underlies the uncertainty in gradient sensing. Based on the above model, below we shall derive the best achievable limit for the sensing uncertainty.

Under the chemical gradient field $\mathcal{G}(\mathbf{x}; p,\phi)$, the log likelihood for observing a specific configuration of cell-surface receptor states is:
\begin{equation}\label{eq:likelihood}
\ln\mathcal{L}(p,\phi)  =  \ln \prod_{m=1}^MP_m^{k_m}(1-P_m)^{N_m-k_m}  \ ,
\end{equation}
where $k_m$ is the number of active receptor clusters in aggregate $m$. An efficient unbiased estimator to parameters associated with the gradient field $\mathcal{G}(\mathbf{x}; p,\phi)$ has a variance limited by the optimal Cramer-Rao lower bound (CRLB) that can be determined from the likelihood function of Eq.(\ref{eq:likelihood}). For convenience, define $(\psi_1,\psi_2)\equiv (p,\phi)$. We can calculate the Fisher information matrix~\cite{kay1993fundamentals}:
\begin{equation}
 [{\bm I}]_{ij}\equiv\left\langle\frac{\partial\ln\mathcal{L}}{\partial\psi_i}\frac{\partial\ln
\mathcal{L}}{\partial\psi_j}\right\rangle, \ \ i, j=1, 2 \ ,
\end{equation}
where the expectation $\langle\cdot\rangle$ is taken with respect to the binomial distribution
$f(k_m;P_m,N_m)={N_m \choose k_m}P_m^{k_m}(1-P_m)^{N_m-k_m}$. We then have
\begin{eqnarray}
[{\bm I}]_{ij} & = & \left\langle\sum_{m=1}^M\sum_{l=1}^M\frac{k_m-N_mP_m}{P_m(1-P_m)}\frac{\partial P_m}{\partial\psi_i}\frac{k_l-N_lP_l}{P_l(1-P_l)}\frac{\partial P_l}{\partial\psi_j}\right\rangle \notag\\
& = & \sum_{m=1}^M\sum_{l=1}^M\frac{\left\langle(k_m-N_mP_m)(k_l-N_lP_l)\right\rangle}{P_m(1-P_m)P_l(1-P_l)}\frac{\partial P_m}{\partial\psi_i}\frac{\partial P_l}{\partial\psi_j} \notag \ .
\end{eqnarray}
Notice that the covariance $\left\langle(k_m-N_mP_m)(k_l-N_lP_l)\right\rangle=0$ when $m\ne l$ due to independence of receptor aggregates. For $m=l$, the variance of $k_m$ is $\left\langle(k_m-N_mP_m)^2\right\rangle=N_mP_m(1-P_m)$. Therefore, the Fisher information matrix is:
\begin{equation}\label{eq:fisher}
[{\bm I}]_{ij}=\sum_{m=1}^M\frac{N_m\left(\partial P_m/\partial c_m\right)^2}{P_m(1-P_m)}\frac{\partial c_m}{\partial\psi_i}\frac{\partial c_m}{\partial\psi_j} \ ,
\end{equation}
where one can verify that
\begin{equation}
\frac{\partial P_m}{\partial c_m}=-nP_m(1-P_m)\frac{1-\alpha}{(1+c_m)(1+\alpha c_m)} \ .
\end{equation}
We define the coefficient
\begin{equation}\label{eq:omega}
\omega_m\equiv\frac{N_m\left(\partial P_m/\partial c_m\right)^2}{P_m(1-P_m)}=\frac{N_mn^2P_m(1-P_m)(1-\alpha)^2}{(1+c_m)^2(1+\alpha c_m)^2} \ ,
\end{equation}
which is a function of local chemical concentration $c_m$, size of the receptor aggregate $N_m$, the strength of receptor coupling $n$ and receptor affinities to the ligand $K_{\rm on}$ and $K_{\rm off}$. We first notice that for the single receptor aggregate $m$ sensing its local chemical concentration $c_m$, the Cramer-Rao lower bound of the sensing variance is given as:
\begin{equation}\label{eq:cm}
\sigma_{c_m}^2=\frac{1}{\omega_m}= \frac{(1+c_m)^2(1+\alpha c_m)^2}{N_mn^2P_m(1-P_m)(1-\alpha)^2} \ .
\end{equation}
The noise-to-signal ratio is determined by $\sigma_{c_m}^2/c_m^2$.

For chemical gradient detection by all cell-surface receptor aggregates, the sensing limit to steepness $p$ or direction $\phi$ is set by the CRLB's (diagonal entries of the inverted Fisher information matrix):
\begin{equation}\label{eq:inv}
\sigma_p^2=[{\bm I}^{-1}]_{11},  \ \ \ \sigma_{\phi}^2=[{\bm I}^{-1}]_{22} \ .
\end{equation}

To obtain analytical results, we assume that the cell resides in a linear gradient field. Across a typical cell size ({\it e.g.}, about 10 $\mu$m for {\it Dictyostelium} and 1 $\mu$m for {\it E. coli}), the linear gradient is a reasonable approximation. $c_m$ at the polar coordinate $(r_m,\theta_m)$ is: 
\begin{equation}\label{eq:lin}
c_m=c_0[1+\beta_mp\cos(\theta_m-\phi)] \ , 
\end{equation}
where $c_0$ is the background concentration at the origin and the coefficient $\beta_m\equiv r_m/r_0$ is the distance from aggregate $m$ to the origin normalized by the reference distance $r_0$. We have
\begin{equation}
 \frac{\partial c_m}{\partial p}=c_0\beta_m\cos(\theta_m-\phi), \frac{\partial c_m}{\partial\phi}= c_0\beta_mp\sin(\theta_m-\phi) \ .
\end{equation}
One can now obtain the Fisher information matrix:
\begin{equation}\label{eq:flinear}
{\bm I}=c_0^2\sum_{m=1}^M\omega_m\beta_m^2\left[\begin{array}{cc}\cos^2(\theta_m-\phi) & \displaystyle\frac{p\sin2(\theta_m-\phi)}{2}\\ \displaystyle\frac{p\sin2(\theta_m-\phi)}{2} & p^2\sin^2(\theta_m-\phi) \end{array}\right] \ .
\end{equation}
In all cases, $\sigma_p^2$ and $\sigma_{\phi}^2$ can be numerically computed, and it is possible to obtain their analytical solutions under special cell geometry and surface layout of receptor aggregates. Here we specialize to a circular cell of radius $r$ and designate the origin as the cell center such that $\beta_m=\beta=r/r_0$, for all $m$. For a shallow gradient across the cell length ($p\beta\ll 1$), we approximate $\omega_m\approx\omega_0$ for all receptor aggregates. $\omega_0$ is calculated by Eq.~(\ref{eq:omega}) evaluated at concentration $c_0$, which can be considered as the average ligand concentration at the cell location (assigned to the cell center). For an equidistant layout of identically-sized receptor aggregates over the cell surface (an isotropic distribution), we obtain the CRLB's for uncertainties in sensing $p$ and $\phi$~\footnote{For an equidistant layout of receptor aggregates ($M\ge 3$), $\sum_{m=1}^M\cos^2(\theta_m-\phi)=\sum_{m=1}^M\sin^2(\theta_m-\phi)=M/2$ and $\sum_{m=1}^M\sin2(\theta_m-\phi)=0$.}:
\begin{equation}\label{eq:limi}
\sigma_p^2=\frac{2(1+c_0)^2(1+\alpha c_0)^2}{N_{\rm tot}\beta^2\xi_n(1-\alpha)^2c_0^2}, \ \ \sigma_\phi^2=\frac{\sigma_p^2}{p^2} \ .
\end{equation}
The factor $\xi_n\equiv nP_0(1-P_0)$ is related to receptor cooperativity $n$, where $P_0$ is calculated by Eq.~(\ref{eq:pm}) at $c_0$. The above limits in Eq.(\ref{eq:limi}) have a few properties: (i) Both $\sigma_p^2$ and $\sigma_\phi^2$ are independent of the number of receptor aggregates $M$ and are insensitive to the gradient direction $\phi$. (ii) Directional sensing improves ($\sigma_{\phi}^2$ decreases) as the steepness $p$ increases. Fortuitously, $\sigma_\phi^2$ is the noise-to-signal ratio of steepness sensing. (iii) The receptor system loses sensing capability when ligand binding does not differentiate the two receptor states ($\alpha\approx 1$). (iv) $\sigma_p^2$ is very sensitive to the cell size via the term $N_{\rm tot}\beta$, where $N_{\rm tot}$ may vary with the cell size by a certain relationship and therefore $\sigma_p^2$ could scale more strongly than the inverse of normalized cell radius $\beta^2$. Properties (iii) and (iv) hold even though the layout of receptor aggregates on the cell surface is anisotropic.

\subsection{Extension to a three-dimensional cell model}
The 2-dimensional model can be extended to a 3-dimensional cell to approach a more geometrically realistic scenario. In the spherical coordinate system, a point has its spatial coordinate $(\rho,\vartheta,\varphi)$, where $\rho$ is the radial length, $\vartheta$ and $\varphi$ are the polar angle and the azimuthal angle. Assume a linear chemical gradient with a steepness $p$ and a direction along the unit vector $\mathbf{g}\equiv[\cos\vartheta_g,\sin\vartheta_g\cos\varphi_g,\sin\vartheta_g\sin\varphi_g]$. For a receptor aggregate $m$ at cell-surface location $A_m$, the vector along the direction from origin $O$ at the cell center to $A_m$, $OA_m$, is $\mathbf{a}_m\equiv r_m[\cos\vartheta_m,\sin\vartheta_m\cos\varphi_m,\sin\vartheta_m\sin\varphi_m]$. Concentration $c_m$ can be calculated as:
\begin{equation}
c_m = c_0(1+ \frac{p}{r_0} \mathbf{a}_m\cdot \mathbf{g}) = c_0(1+\beta_mp\cos\gamma_m) \ ,
\end{equation}
where $\gamma_m$ is the angle between $\mathbf{a}_m$ and $\mathbf{g}$. The dot product $\mathbf{a}_m\cdot\mathbf{g}=r_m\cos\gamma_m$ projects $\mathbf{a}_m$ onto the gradient direction $\mathbf{g}$, with
$\cos\gamma_m=\sin\vartheta_m\sin\vartheta_g\cos(\varphi_m-\varphi_g)+\cos\vartheta_m\cos\vartheta_g$.
We have:
\begin{eqnarray}
\frac{\partial c_m}{\partial p} & = & c_0\beta_m\cos\gamma_m \\
\frac{\partial c_m}{\partial\vartheta_g} & = & c_0\beta_mp( \sin\vartheta_m\cos\vartheta_g\cos(\varphi_m-\varphi_g) \notag \\
& & -\cos\vartheta_m\sin\vartheta_g) \\
\frac{\partial c_m}{\partial\varphi_g} & = & c_0\beta_mp \sin\vartheta_m\sin\vartheta_g\sin(\varphi_m-\varphi_g)\ .
\end{eqnarray}
The Fisher information matrix $\mathbf{I}$ can be constructed by Eq.(\ref{eq:fisher}), with the coefficient $\omega_m$ obtained by Eq.(\ref{eq:omega}). Analytical formula like Eq.(\ref{eq:limi}) can be obtained under special cell geometry and receptor aggregate layout on the cell surface. We will present our results for a 2D cell and expect that general conclusions are applicable to a 3D cell.

\begin{figure}
\centering
\includegraphics[scale=0.3]{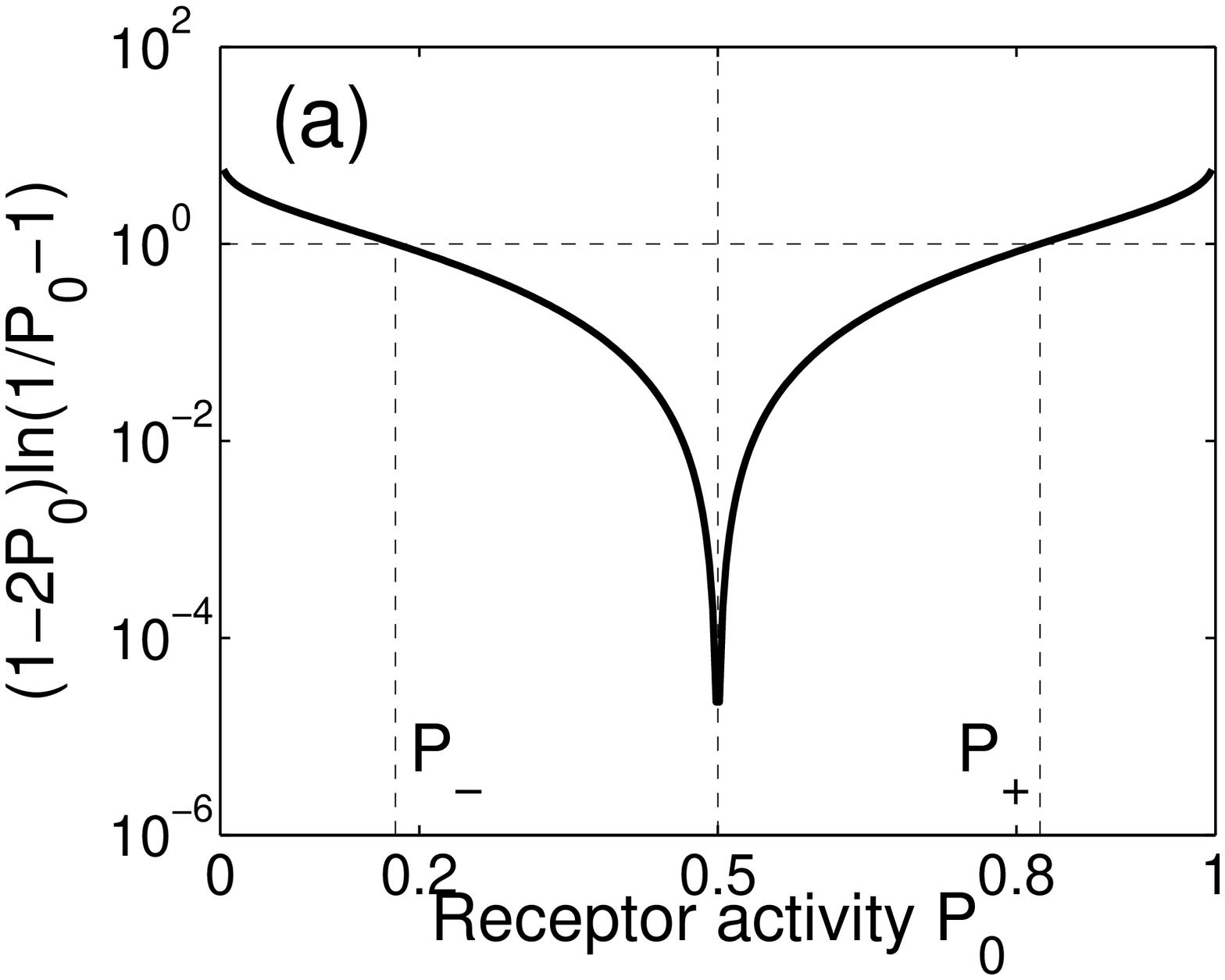}
\includegraphics[scale=0.3]{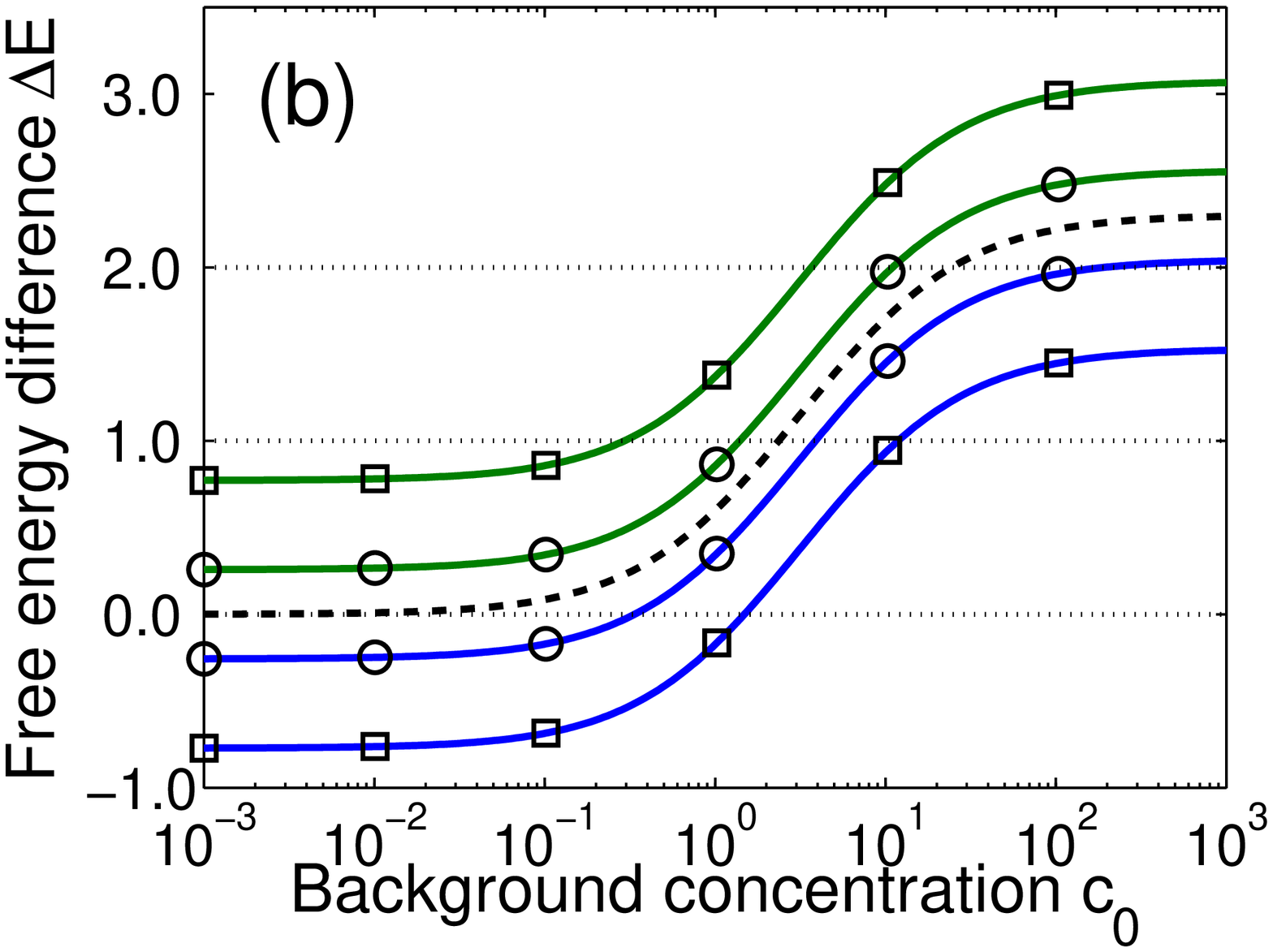}
\caption{\label{fig:constr} (color online). (a). Function $(1-2P_0)\ln(1/P_0-1)$. Inequality Eq. (\ref{eq:p0}) holds when $P_0\in (P_-,P_+)$. (b) Solid curves show $\Delta E$ as a function of $c_0$ at $\alpha=0.1$, $\Delta E=\ln(\frac{1+c_0}{1+\alpha c_0})-\ln(1/P_+-1)/n$ (lower two curves) and $\Delta E=\ln(\frac{1+c_0}{1+\alpha c_0})-\ln(1/P_--1)/n$ (upper two curves), for two cooperativity levels: $n=2$ ($\scriptscriptstyle\square$) and $n=6$ ($\circ$). Upper and lower curves of same $n$ enclose the regime that receptor cooperativity (up to $n$) helps to reduce the variance $\sigma_p^2$. The beneficial ranges of $c_0$ for three different $\Delta E$ values (2.0, 1.0, and 0) are marked (by dotted lines spanning boundaries specified by Eq. (\ref{eq:range})). The dashed line indicates the $\Delta E$ adjusted by precise adaptation, in which the adapted level of receptor activity is a constant $P_a=1/2$.}
\end{figure}

\section{Results}
Here we investigate the role of receptor cooperativity on the gradient sensing limit described by Eq.(\ref{eq:limi}). Under a fixed number of receptors ($N_{\rm tot}$), the effect of cooperativity can be detected as the direction of change in $\xi_n$ with regard to the receptor cluster size $n$:
\begin{equation}
\frac{\partial\xi_n}{\partial n}=P_0(1-P_0)\left[1-n(1-2P_0)\left(\ln\frac{1+c_0}{1+\alpha c_0}-\Delta E\right)\right] \ .
\end{equation}
Beneficial effect of receptor cooperativity to the sensing limit requires $\partial\xi_n/\partial n>0$, which together with $P_0=[1+(e^{-\Delta E}\frac{1+c_0}{1+\alpha c_0})^n]^{-1}$ implies the constraint:
\begin{equation}\label{eq:p0}
(1-2P_0)\ln\left(\frac{1}{P_0}-1\right)<1,
\end{equation}
or equivalently,
\begin{equation}\label{eq:pmp}
P_-<P_0<P_+ \ ,
\end{equation}
where $P_-\approx 0.176$ and $P_+=1-P_-\approx 0.824$ are the two solutions to: $(1-2P_0)\ln(1/P_0-1)=1$ (see Fig.\ref{fig:constr}(a) for illustration). The feasible range of $\Delta E$, $\alpha$ and the background concentration $c_0$ is:
\begin{equation}\label{eq:range}
\frac{\ln\left(1/P_+-1\right)}{n}<\ln\left(\frac{1+c_0}{1+\alpha c_0}\right)-\Delta E<\frac{\ln\left(1/P_--1\right)}{n} \ ,
\end{equation}
within which receptor cooperativity up to the magnitude of $n$ can improve the accuracy of chemical sensing. Fig.~\ref{fig:constr}(b) shows that the beneficial region of free energy difference $\Delta E$ is confined in a banded area from around 0 at low ligand concentration to around $\ln(1/\alpha)$ at high ligand concentration. The bandwidth of $\Delta E$ is a constant $\frac{2}{n}\ln(P_+/P_-)$, limited by the strength of receptor cooperativity $n$. A stronger cooperativity results in a narrower band. For each given $\Delta E$, the beneficial range of background concentration $c_0$ is determined by Eq.~(\ref{eq:range}) (see Fig.\ref{fig:constr}(b) for illustration). Results in Fig.\ref{fig:constr}(b) were obtained at $\alpha<1$, in which ligand binding favors the ``off" state of the receptor. The results at $\alpha>1$ (not shown) are clearly symmetric to those at $\alpha<1$. We note that the above derivation is under the assumption that the receptor system has a fixed total number of receptors $N_{\rm tot}$. The general conclusion also applies to a system that has fixed numbers of receptor clusters ($N_m$) in receptor aggregates~\footnote{In this scenario, the factor related cooperativity is $\xi_n=n^2P_0(1-P_0)$. $\partial\xi_n/\partial n>0$ implies that $(1-2P_0)\ln(1/P_0-1)<2$, or, $P_-<P_0<P_+$, where $P_- \approx 0.083$ and $P_+\approx 0.917$.}.

In the following, we analyze the influence of receptor cooperativity on the chemical sensing limit under the context of receptors with or without adaptation.

\subsection{Non-adaptative receptors}
\begin{figure}
\centering
\includegraphics[scale=0.22]{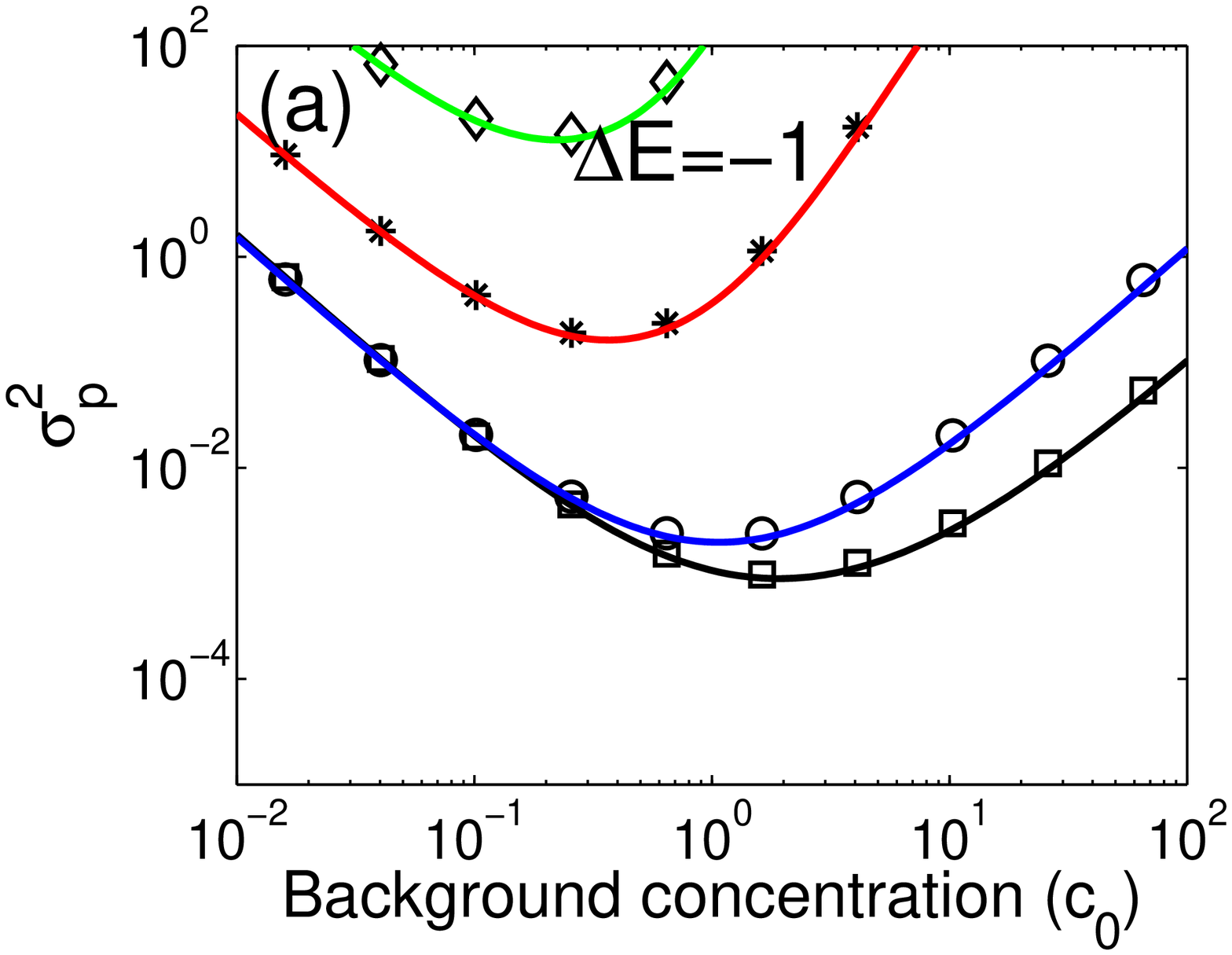}
\includegraphics[scale=0.22]{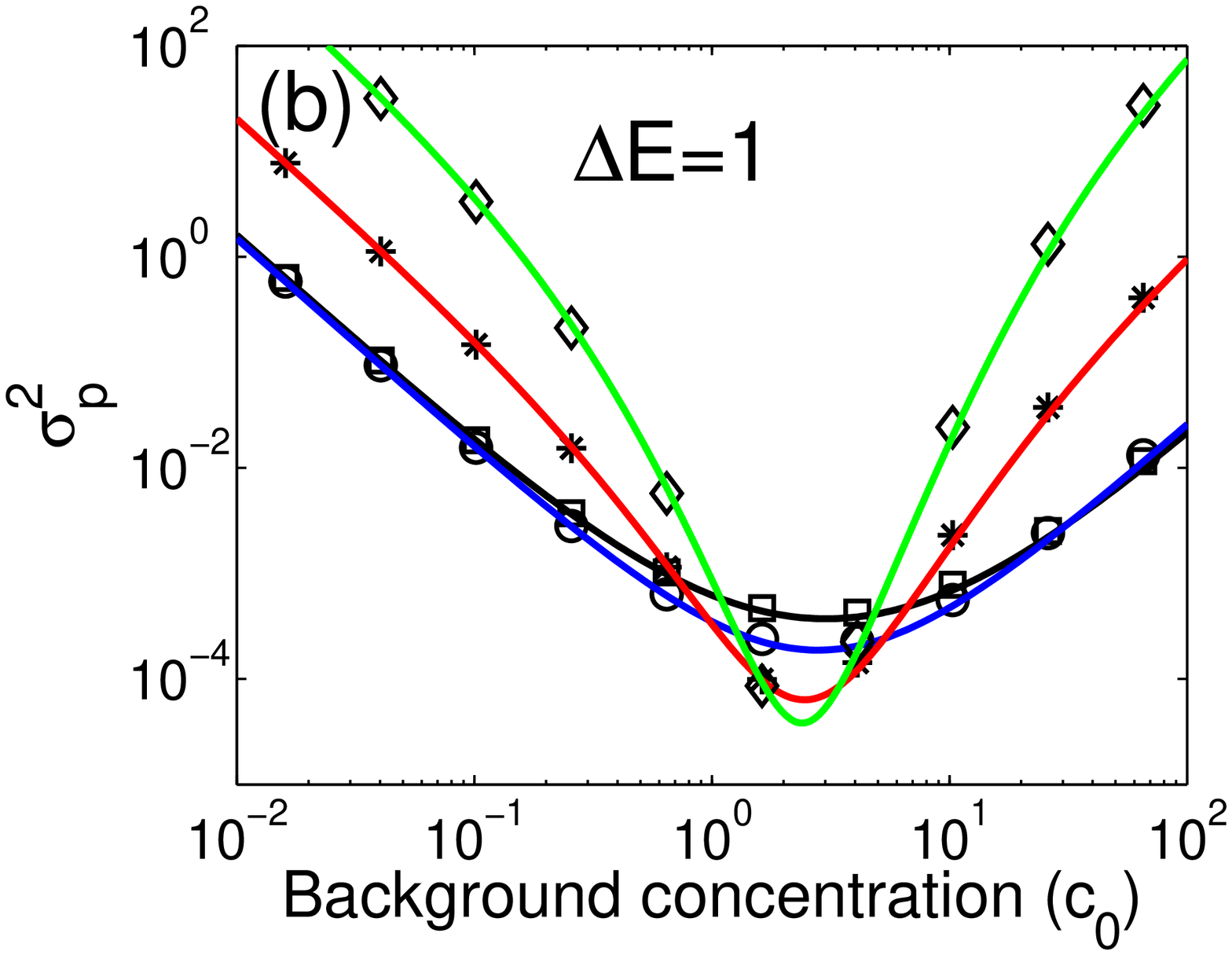}

\includegraphics[scale=0.22]{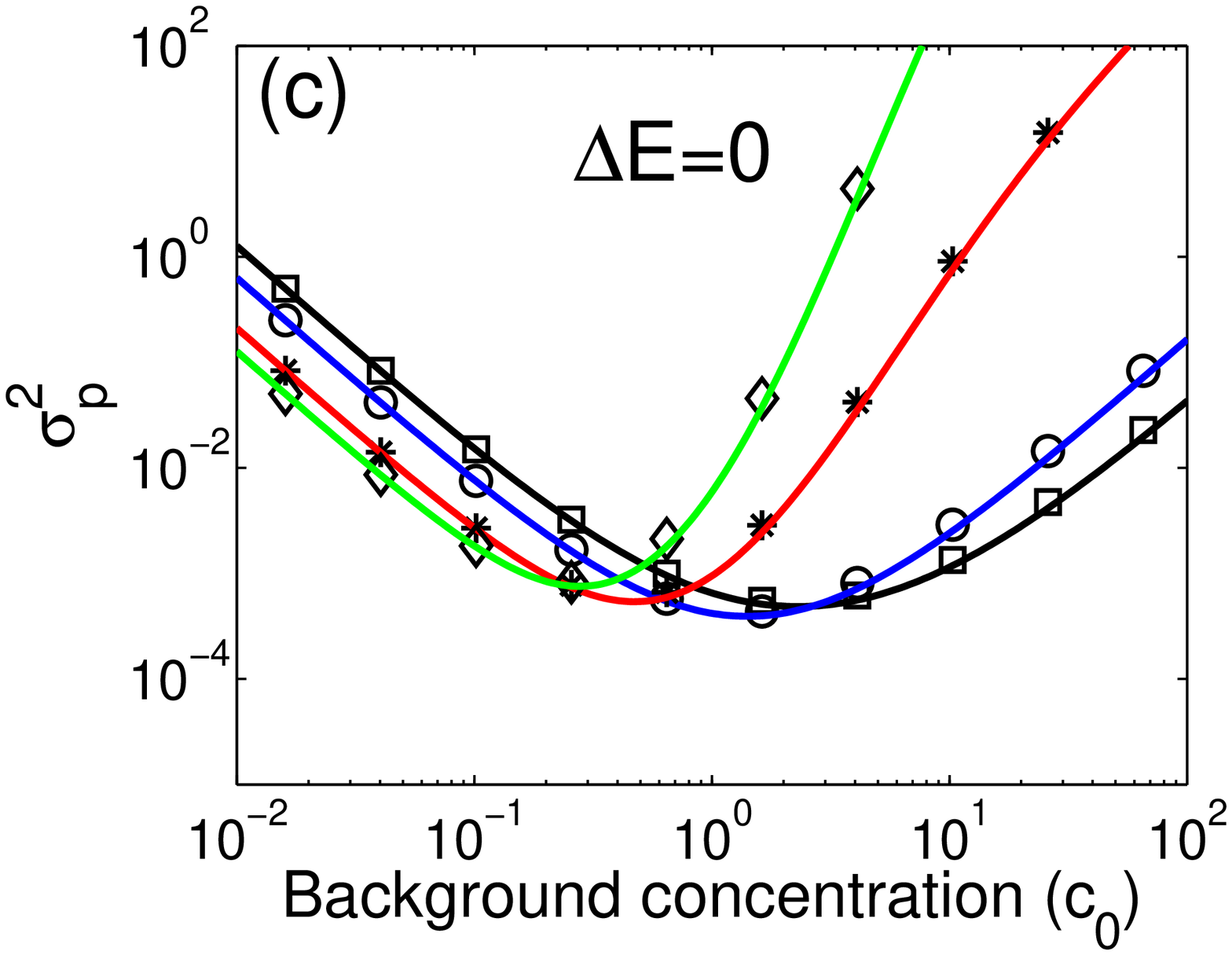}
\includegraphics[scale=0.22]{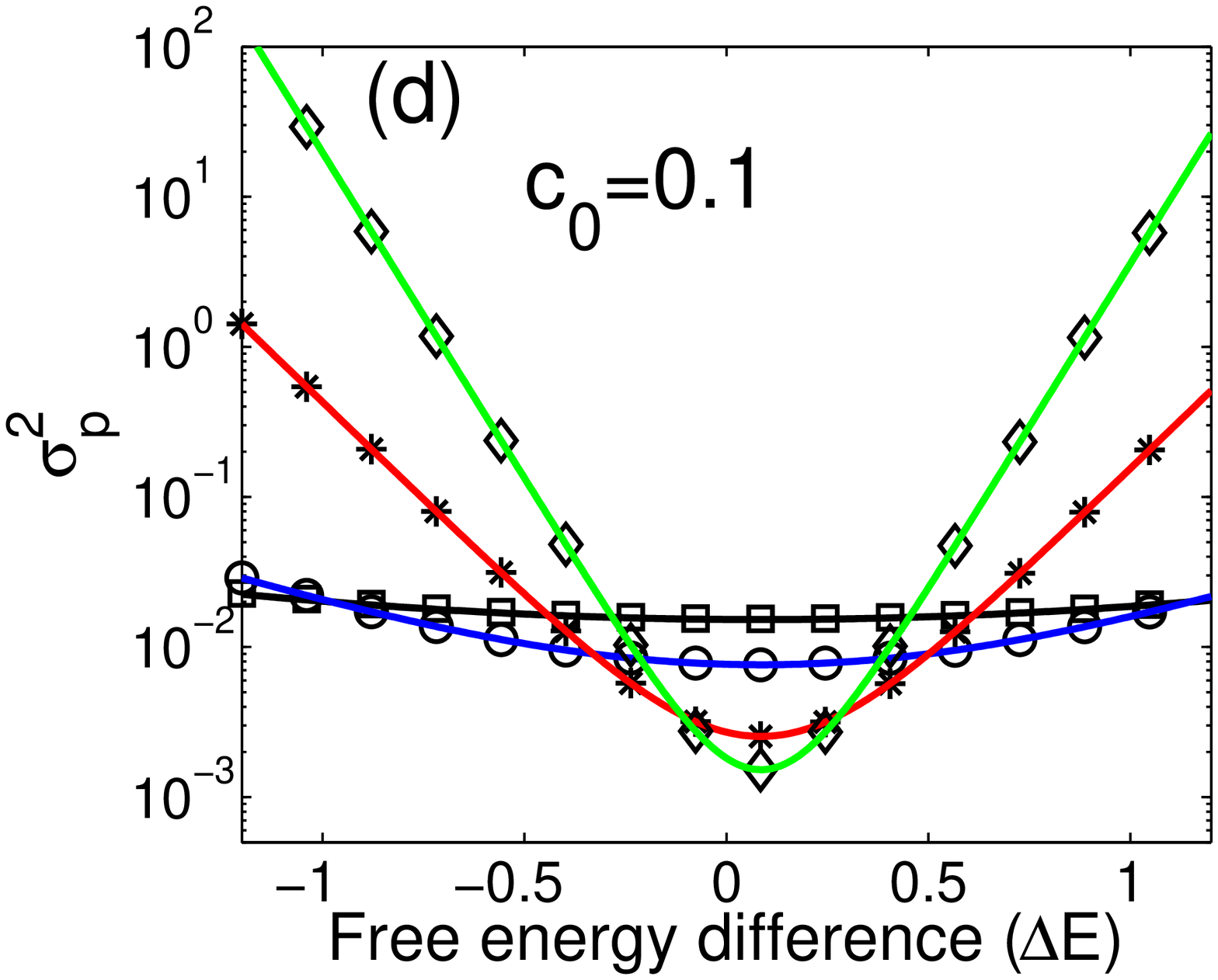}
\caption{\label{fig:coop} (color online). $\sigma_p^2$ under varying background concentrations. Receptor aggregates have an equidistant layout on the cell surface. (a)-(d) were plotted with different $\Delta E$ (values indicated in plots) under different cluster size: $n=1$ (non-cooperative), `$\scriptscriptstyle\square$'; 2 (receptor dimer), `$\circ$'; 6, `$\ast$' (corresponding to ``trimer-of-dimers" in bacteria); and 10, `$\diamond$' (corresponding to an estimate by Ref.~\cite{keymer2006chemosensing}). $\sigma_p^2$ were calculated by Eq.~(\ref{eq:inv}) (markers) and by Eq.~(\ref{eq:limi}) (solid lines) with $N_{\rm tot}=80,000, M=3, p=0.1, \phi=0$, $\alpha=0.1$ and $\beta=1$. $\sigma_\phi^2$ behaves similarly and is not shown.}
\end{figure}

Without receptor adaptation, the equilibrium level of receptor activity changes with the ligand concentration $c_0$ according to Eq.~(\ref{eq:pm}), where free energy difference $\Delta E$ and binding affinities $K_{\rm on}$ and  $K_{\rm off}$ are unmodulated by the receptor activity.  Fig.~\ref{fig:coop} shows that in all cases $\sigma_p^2$ assumes a valley-shaped relationship with the background concentration $c_0$. $\sigma_p^2$ attains a minimum at a $c_0$ that satisfies the condition $d\sigma_p^2/dc_0=0$, {\it i.e.}:
\begin{equation}\label{eq:optimc0}
2\alpha c_0^2+n(1-\alpha)(1-2P_0)c_0-2=0 \ .
\end{equation}

At the regime $\Delta E<0$, where an unliganded receptor is biased to the ``off" state, increasing receptor cooperativity always reduces sensing accuracy [Fig.~\ref{fig:coop}(a)] when ligand binding as well favors the ``off" state ({\it i.e.}, $\alpha<1$). According to Fig.~\ref{fig:constr}(b), $\Delta E=-1$ is out of the beneficial band at any background concentration $c_0$. Intuitively, in this case, receptor arrays have limited capacity (most receptors are already ``off" even before the introduction of ligands) to respond to ligand binding by further switching off receptor activity.

At the regime $\Delta E>0$, receptor cooperativity reduces $\sigma_p^2$ within a concentration range around the optimal $c_0$ [Fig.~\ref{fig:coop}(b)]. Such advantage is confined in a narrowed dynamic range around intermediate background concentrations [the stronger the cooperativity, the shorter the range of improvement] instead of being effective at lower concentrations, a desirable region for chemotactic responses. The results shown in Fig.~\ref{fig:coop}(b) approximate those by the Ising-type model of Hu {\it et al.}~\cite{hu2010physical} in which a receptor maintains the ``on" state in the absence of ligand and is only switched off by ligand binding. In fact, the model by Hu {\it et al.}~\cite{hu2010physical} can be considered as a limiting case to our model, by requiring $e^{-\Delta E}\ll 1$ and $\alpha\ll 1$.

By contrast, receptor cooperativity improves sensing accuracy below the optimal $c_0$ near $\Delta E\approx 0$ [Fig.~\ref{fig:coop}(c)]. Fig.~\ref{fig:coop}(d) shows that at low background concentration ($c_0=0.1$) the system achieves optimal sensing accuracy at free energy difference $\Delta E\approx 0$ of any cooperativity level and receptor cooperativity further decreases $\sigma_p^2$ (an order of magnitude improvement at $n=10$ over independent receptors). We can relate this result to {\it E. coli}, in which $c_0=0.1$ corresponds to a background concentration of 2 nM MeAsp, a threshold level, at $K_{\rm off}=20$ nM and cooperativity predicted about $n=10$ as in Ref.~\cite{keymer2006chemosensing}.

\subsection{Adaptative receptors}
Near precise adaptation was found in bacteria such as {\it E. coli}~\cite{berg1972chemotaxis}. A ligand concentration change triggers a transient response in receptor activity followed by a slow decay back to the steady state about the prestimulus level. In other words, receptor adaptation desensitizes the steady-state activity to ligand concentration, allowing the cell to be responsive to environmental signals within a wide dynamic range. Here we show that adaptation also conditions the receptor system to allow cooperativity to improve signal-to-noise ratio in chemical sensing under a wide range of background concentrations.

We assume that the receptor system is adapted to its chemical environment before an onset of change in ligand concentration. We also assume that the adaptation machinery adjusts the free energy difference $\Delta E$ between the two states of the receptor and keeps binding affinities $K_{\rm on}$ and $K_{\rm off}$ invariant. For the example of {\it E. coli}, receptor activity-controlled receptor methylation and demethylation act as a feedback mechanism that can adjust the free energy gap $\Delta E$. For precise adaptation, equilibrium receptor activity is tuned to be a constant, $P_a$, independent of the background concentration. By Eq.~(\ref{eq:pm}), at aggregate $m$, we have the adjusted free energy difference:
\begin{equation}
\Delta E=\ln\frac{1+c_m}{1+\alpha c_m}-\frac{1}{n}\ln\left(\frac{1}{P_a}-1\right) \ .
\end{equation}
Concentration changes around $c_m$ induces transient receptor activity $P_m$ away from $P_a$, which is described by Eq.(\ref{eq:pm}) with the above adjusted $\Delta E$.

The optimal sensing achieves at the half-activation level $P_a=1/2$ that maximizes $\xi_n$, where $\sigma_p^2=8(1+c_0)^2(1+\alpha c_0)^2/(N_{\rm tot}n\beta^2(1-\alpha)^2c_0^2)$. $\sigma_p^2$ attains a minimum of $\sigma_{p,\rm min}^2=8(\alpha^{1/2}+1)^4/(N_{\rm tot}n\beta^2(1-\alpha)^2)$ at the background concentration $c_0=\alpha^{-1/2}$ according to Eq.~(\ref{eq:optimc0}) [{\it i.e.}, $[L]_0=(K_{\rm on}K_{\rm off})^{1/2}$ is the geometric mean of dissociation constants]. Fig.\ref{fig:constr}(b) shows (dashed line) the adjusted $\Delta{E}$ to adapted receptor activity $P_a=1/2$ as a function of ligand concentration. The adaptation tunes $\Delta{E}$ into the middle of the beneficial region band and allows the receptor cooperativity to improve sensing of small concentration changes. 

The adaptation mechanism may also likely maintain a constant steady-state activity by adjusting ligand binding affinities $K_{\rm on}$ and $K_{\rm off}$, or adjusting all three thermodynamics parameters altogether~\cite{mello2007effects}. How adaptation tunes these parameters altogether remains unclear even for the well-studied {\it E. coli} chemoreceptors. Nonetheless, as our model shows, as long as a cell achieves an adaptation level (precise or imprecise) that satisfies $P_-<P_a<P_+$, receptor cooperativity is advantageous for sensing small changes in chemical concentration or gradient. It has been suggested a precise adaptation of $P_a\approx 1/3$ in {\it E. coli}~\cite{clausznitzer2010chemotactic} in the range specified by Eq.~(\ref{eq:pmp}), supporting the idea that receptor adaptation and cooperativity act in concert to improve chemical sensing. 

\subsection{Effects of anisotropy in the layout of receptor aggregates}
The analytical formula in Eq.~(\ref{eq:limi}) do not apply to the scenario where receptor aggregates anisotropically locate on the cell surface. As we will show, such an anisotropy may cause substantial dependence of $\sigma_p^2$ and $\sigma_\phi^2$ on the gradient direction $\phi$, the number and locations of receptor aggregates. We define a metric, $\mathcal{R}_e$, to quantify the layout anisotropy:
\begin{equation}\label{eq:re}
 \mathcal{R}_e=\frac{\sum_{i=1}^{M}|\theta_{i+1}-\theta_i-2\pi/M|}{4\pi(M-1)/M} ,
\end{equation}
which is a normalized mean angular dispersion between two immediately adjacent aggregates from the average angular distance, $2\pi/M$. Without loss of generality, the angles of receptor aggregates are labeled in an order such that $\theta_{i+1}\ge\theta_i$, for $i=1,..,M$, with the periodic boundary condition: $\theta_{M+1}\equiv\theta_1+2\pi$ (see Fig.\ref{fig:scheme}). $\mathcal{R}_e$ ranges from 0 for the isotropic case to 1 when all receptor aggregates gather at a single location.

Each effective receptor aggregate acts as a local concentration sensor, and a pair of aggregates can detect the projection of the gradient onto the direction connecting the two aggregates. Two independent projections are needed in minimum to reconstruct the gradient. In our model, the minimal number of effective cell-surface receptor aggregates is $M=2$ because of the implicit assumption that one parameter of the linear gradient, the background concentration $c_0$, is known. This assumption is justified by considering that the cell has a memory for a recent history of its chemical environment. 

\begin{figure}[t]
\centering
\includegraphics[scale=0.3]{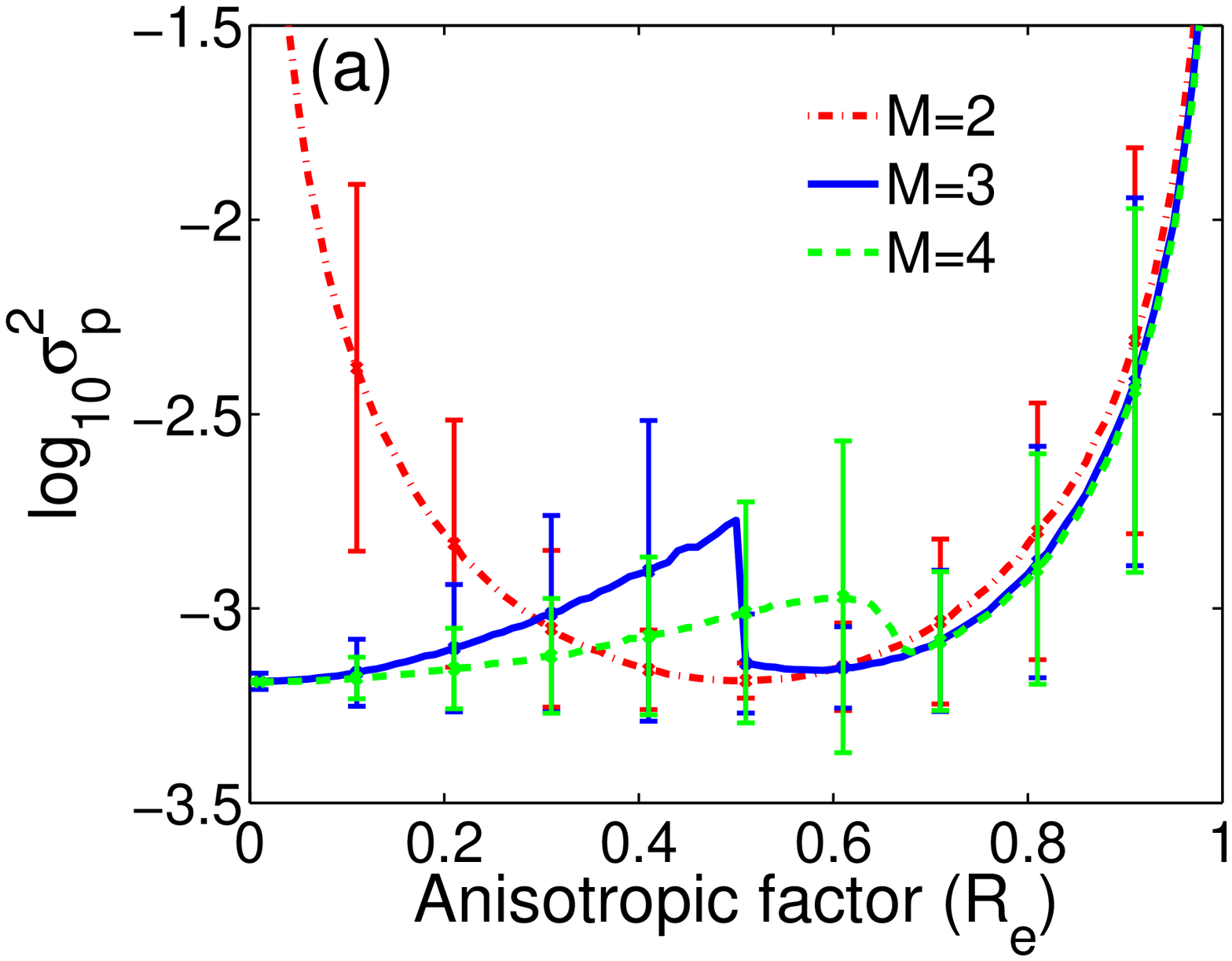}

\includegraphics[scale=0.3]{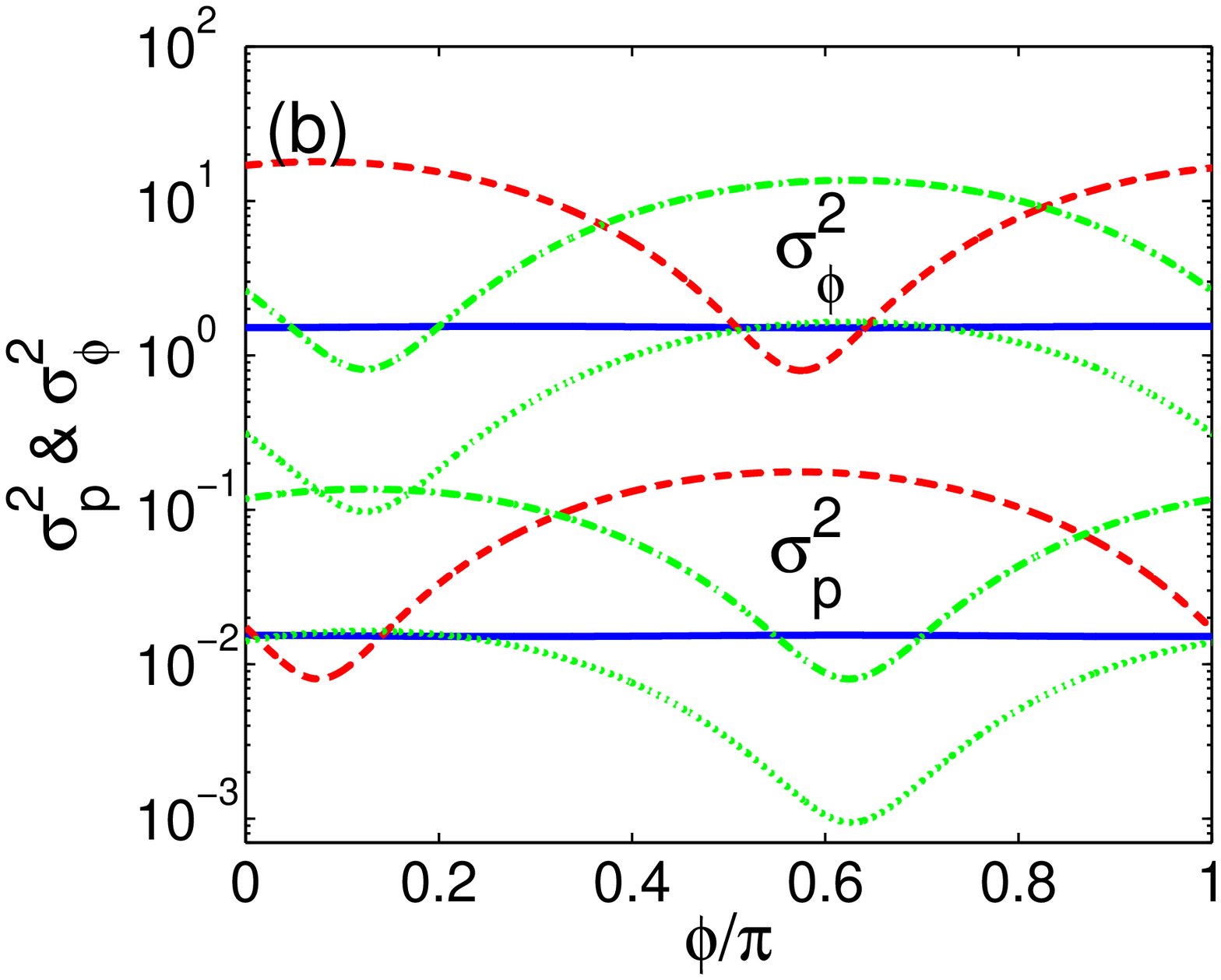}
\caption{\label{fig:fig3} (color online). Sensing limit under varying receptor aggregate layout and cell orientation (by non-adaptor receptors). (a) $\sigma_p^2$ vs. $\mathcal{R}_e$ ($n=1$) at $M=2$ (dashed dot), 3 (solid) and 4 (dashed). $R_e$ was partitioned into 100 equispaced bins within the interval $(0,1)$ and $\sigma_p^2$ was the geometric mean over 10,000 instances in each bin. Standard errors around the means are shown at several $R_e$ values. (b) $\sigma_p^2$ (lower 4 curves) and $\sigma_\phi^2$ (upper 4 curves), equidistance layout (solid) vs. two anisotropic layouts without receptor cooperativity: (i) $\mathcal{R}_e=0.38$, $\theta_1=0.156\pi$, $\theta_2=1.08\pi$ and $\theta_3=1.99\pi$ (dashed) and (ii) $\mathcal{R}_e=0.49$, $\theta_1=0.513\pi$, $\theta_2=1.66\pi$ and $\theta_3=1.68\pi$ (dashed dot), or with receptor coupling of size $n=10$ for layout (ii) (dotted). Cell rotation is simulated by relatively varying $\phi$ from 0 to $\pi$. $\sigma_p^2$ and $\sigma_\phi^2$ were calculated by Eq.~(\ref{eq:inv}) with parameter values $N_{\rm tot}=80,000$, $M=3$, $p=0.1$, $c_0=0.1$, $\alpha=0.1$, $\beta=1$,  and $\Delta E=0$.}
\end{figure}

A sensing singularity occurs when the cell (1) has only one effective receptor aggregate because two aggregates locate too close to each other, or (2) has two aggregates lying on the opposite side across the cell center such that only one independent projection of the chemical gradient can be detected. For such a special layout, the Fisher information matrix of Eq.~(\ref{eq:flinear}) becomes degenerate and the cell cannot resolve either gradient steepness $p$ or the direction $\phi$. For $M=2$, $\sigma_p^2$ diverges at either end, $\mathcal{R}_e=0$ or $1$, where the system has a single independent sensor [Fig.~\ref{fig:fig3}(a)]. For $M=3$ or $M=4$, $\sigma_p^2$ diverges at $\mathcal{R}_e=1$ and a singularity of the sensing limit may also happen at $\mathcal{R}_e=1/2$ or 2/3, respectively. Notice that, as the number of receptor aggregates $M$ increases, the extent of singularity decreases because receptor aggregate layouts with the corresponding value of $\mathcal{R}_e$ become less likely to be degenerate by chance. 

\begin{figure}[t]
\centering
\includegraphics[scale=0.25]{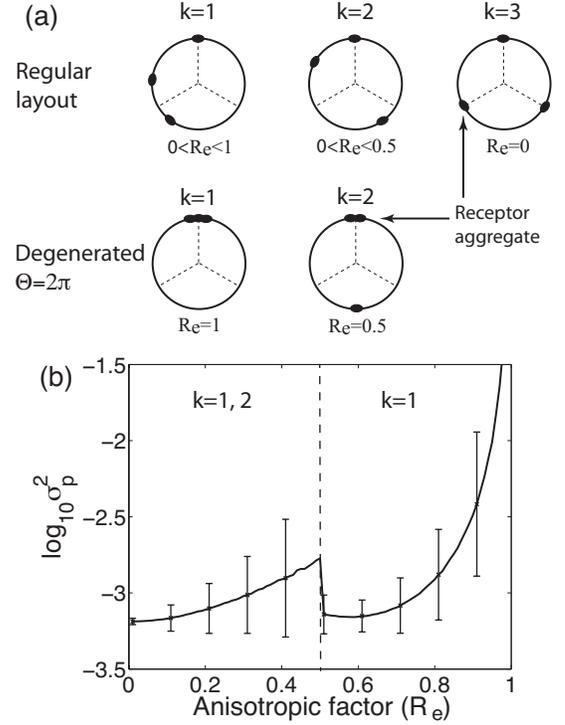}
 \caption{\label{fig:si1} (a) Sample layouts of receptor aggregates ($M=3$). Note that $\mathcal{R}_e=0.5$ corresponds to any scenario of the cell with two effective aggregates on cell surface and only the one shown with two aggregates on the opposite sides across the cell center is degenerate. Dashed lines inside the cell circle give a reference to an equiangular partition. (b) Correspondence between possible $k$ values and $\mathcal{R}_e$ and their relationship with $\sigma_p^2$ (shown as the geometric mean with s.e.m.). \label{fig:s1}}
\end{figure}

In general, one can verify that $\sigma_p^2$ may become singular at $\mathcal{R}_e=(M-2)/(M-1)$ and $\mathcal{R}_e=1$ because some receptor aggregates locate so close to each other that they effectively sense a same local ligand concentration. Suppose among all immediately neighboring aggregates along the cell circle there exist $k$ angles that are equal to or greater than the average angle $2\pi/M$ ({\it i.e.}, the rest $M-k$ angles are smaller than $2\pi/M$). Let $\Theta$ be the sum of these $k$ angles. We rewrite the anisotropic factor of Eq. (\ref{eq:re}) as:
\begin{equation}\label{eq:re2}
 \mathcal{R}_e=\frac{\Theta-2\pi k/M}{2\pi (M-1)/M} \ ,
\end{equation}
which changes from 0 to 1 as $\Theta$ ranges from $2\pi k/M$ to $2\pi$. The aggregate layout can be classified into three cases when the value of $k$ changes (see Fig.~\ref{fig:s1}(b) for example layouts when $M=3$). (i) When $k=1$ and $\Theta=2\pi$, $\mathcal{R}_e=1$, the system has a single effective aggregate at one location ($\mathcal{R}_e=1$), and therefore cannot resolve parameters $p$ or $\phi$. The system is non-degenerate as long as $\Theta\ne 2\pi$. (ii) When $k=2$ and $\Theta=2\pi$, $\mathcal{R}_e=(M-2)/(M-1)$, there exist two effective aggregates. In the special case in which these two effective aggregates locate at two opposite sides on the circle across the cell center, the Fisher information matrix degenerates. The system is non-degenerate as long as $\Theta\ne 2\pi$. (iii) When $k\ge 3$, no degeneration appears. As an illustration for $M=3$, Fig.\ref{fig:si1}(b) shows that a singularity in $\sigma_p^2$ may happen at $\mathcal{R}_e=0.5$ (corresponding to the last layout in Fig.~\ref{fig:si1}(a)). However, the majority of layouts with $\mathcal{R}_e=0.5$ are non-degenerate. For instance, all layouts with $k=1$ and $\Theta=4\pi/3$ has $\mathcal{R}_e=0.5$ but none of them is degenerate.

Both $\sigma_p^2$ and $\sigma_{\phi}^2$ are sensitive to the cell orientation under anisotropic aggregate layout [see Fig.~\ref{fig:fig3}(b) for a non-adaptive case]. The effect of receptor cooperativity is preserved under anisotropic receptor aggregate layouts because $\xi_n$ is geometrically independent under a shallow gradient. By changing the cell orientation, improvement in estimating one parameter of $p$ or $\phi$ is gained at the expense of the other. This result coincides with the study by Hu {\it et al.}~\cite{hu2011geometry} who showed that an elliptic cell cannot simultaneously improve limits $\sigma_p^2$ and $\sigma_\phi^2$ by elongating the cell body. As an optimal configuration, the isotropic layout of receptor aggregates improves sensing both $p$ and $\phi$ in most cases and is insensitive to cell orientation as indicated by Eq.~(\ref{eq:limi}) and as shown in Fig.~\ref{fig:fig3}(b). This result recapitulates the one derived by Berg and Purcell~\cite{berg1977physics} who showed that uniformly distributed receptor patches on the membrane maximize independent ligand influx to the cell and thus sensitizes the concentration measurement. However, as shown by Fig.~\ref{fig:fig3}(b), proper cell orientation can improve estimation of one of the parameters ($p$ or $\phi$), which might be a desirable task if sensing one parameter is more critical than the other. For example, accurate directional sensing may be more important than steepness sensing in chemotaxis so that a cell can identify swimming direction more efficiently. For a cell with anisotropic layout of receptor aggregates, routine reorientation is required to improve sensing performance (e.g., tumbling of {\it E. coli}). 

The above results of minimal number and optimal layout of receptor aggregates were obtained under the assumption that the background ligand concentration $c_0$ is known. One can also consider $c_0$ as an extra parameter to be estimated by the cell. In this case, merely by counting the number of parameters ($p$, $\phi$ and $c_0$) the cell needs in minimum three independent receptor aggregates as concentration sensors to reliably reconstruct the gradient. To compute the CRLB's, one must also calculate the derivative $\partial c_m/\partial c_0$ and obtain a 3-by-3 Fisher information matrix. The primary results about the sensing limits remain little changed. It is straightforward to verify that under isotropic receptor layout the variances of parameter estimates:
\begin{eqnarray}
\sigma_p^2 & = & \frac{(2+\beta^2p^2)(1+c_0)^2(1+\alpha c_0)^2}{N_{\rm tot}\beta^2\xi_n(1-\alpha)^2c_0^2}, \\ 
\sigma_\phi^2 & = & \frac{\sigma_p^2}{(1+\beta^2p^2/2)p^2}, \\ 
\sigma_{c_0}^2 & = & \frac{(1+c_0)^2(1+\alpha c_0)^2}{N_{\rm tot}\xi_n(1-\alpha)^2}\ .
\end{eqnarray}
Under a shallow gradient ($\beta p\ll1$), $\sigma_p^2$ approaches the one derived in Eq.~(\ref{eq:limi}), while the variance of the directional sensing $\sigma_\phi^2$ is unchanged. The variance $\sigma_{c_0}^2$ is in a similarly form of Eq.~(\ref{eq:cm}) for detecting local concentration $c_m$ by a single receptor aggregate. 

\section{Discussion and conclusion}

It is well-known that adaptation brings the receptor activity back to pre stimulus level, allowing the system to remain sensitive to chemical changes in the future. This mechanism enables the system to respond within a wide dynamic range of chemical concentration. Here we examine the role of adaptation in terms of chemical sensing accuracy, especially when presented with small signals or small changes in the signal, where stochastic noise may become overwhelming. Our study showed that receptor cooperativity improves sensing accuracy only within a limited background concentration range if the receptor aggregates do not undergo adaptation. By contrast, receptor adaptation, especially precise or near precise adaptation, maintains the receptor sensory system to operate within the parametric region where receptor cooperativity is beneficial. As to our knowledge, the finding in this work for the first time connects receptor adaptation and cooperativity to noise filtering in chemical sensing. 

The sensing limit of Eq.~(\ref{eq:limi}) was derived from the uncertainty of an instantaneous receptor sampling of a gradient field. The sensing accuracy can be improved by a cell integrating independent receptor state configurations over time~\cite{berg1977physics}. The time interval between two independent measurements is determined by the correlation time $\tau_c$ that accounts for relaxation times of ligand diffusion, ligand-receptor binding~\cite{hu2010physical} and receptor dynamics. Because the amount of time $t$ available for averaging is limited by dynamics of the intracellular signaling circuit, the number of independent measurements is about $\frac{t}{2\tau_c}$~\cite{wang2007quantifying}. Therefore, the temporal averaging reduces the instantaneous limit $\sigma_p^2$ to $\langle\sigma_p^2\rangle_t\approx 2\tau_c\sigma_p^2/t$ when $t\gg 2\tau_c$. Ligand concentration is encoded as receptor occupancy by the ligand-receptor binding and is then transduced into receptor activity by the MWC model. Stochastic dynamics of each of these signal transduction stages contributes additional noise and thus increases the correlation time $\tau_c$, which consequently elevates the sensing uncertainty within the given time frame $t$.

Our study does not consider the temporal dynamics of ligand-receptor binding and receptor state switching. Skoge {\it et al.}~\cite{skoge2011dynamics} recently showed that receptor cooperativity may significantly slow down receptor state switching (described by Glauber dynamics~\cite{glauber1963time} originally developed for studying time dependence of the Ising model) to an extent such that the system cannot effectively relay the information of ligand concentration changes in time [$\tau_c$ becomes comparable or greater than the averaging time $t$]. However, there is still a lack of direct experimental evidence on how receptor coupling quantitatively modulates receptor dynamics. Our results reveal that the benefit provided by receptor coupling to chemical sensing by instantaneous receptor activity configurations may compensate the potentially detrimental effect of receptor activity slowdown, in particular by adapted receptor arrays.

Another possible source of noise at the ligand-receptor interaction level is due to ligand rebinding when a dissociated ligand diffuses back again to bind onto the receptor aggregate before it escapes into the bulk medium. The extent of ligand rebinding and its effects on the accuracy of ligand sensing depends on the size and density of a receptor aggregate~\footnote{The effective rate constants under ligand rebinding are given as: $\tilde{k}_f=k_f/(1+[R]k_f/k_+)$ and $\tilde{k}_r=k_r/(1+[R]k_f/k_+)$, where $k_+=4Da$ is the diffusion limited on rate to a receptor aggregate of size $a$. $[R]k_f/k_+$ is the rebinding/escape ratio that characterizes the degree of influence by ligand rebinding.}. Theory~\cite{shoup1982role,*goldstein1995approximating} and simulation~\cite{andrews2005serial} showed that ligand rebinding does not change the equilibrium occupancy of individual receptors and therefore does not affect the sensing limits obtained from instantaneous measurement. However, significant ligand rebinding does introduce extra fluctuations by decreasing the effective rate constants for ligand association with and dissociation from the receptor, which increases the correlation time $\tau_c$ and thus reduces the number of independent measurements within a fixed averaging time.

Chemoreceptor cooperativity in bacteria was predicted by Ising models and MWC models used for analyzing dose response data in {\it E. coli}. More recently receptor coupling is evident by the hexagonal organization of {\it trimer-of-dimers} of chemoreceptors resolved by electron cryotomography as a conserved architecture in a wide variety of bacteria~\cite{briegel2009universal,*briegel2011activated,*Briegel21022012,*liu2012molecular}. The spatial organization of eukaryotic chemoreceptors remains to be fully resolved, even though higher-order receptor arrays in chemotactic eukaryotes have also been observed~\cite{wadhams2004making}, suggesting possible receptor coupling. Our model of the receptor organization on the cell surface is general and can be parameterized to study systems of coupling or non-coupling receptors with or without adaptation. 

Berg and Purcell~\cite{berg1977physics} showed that uniformly distributed receptor patches over the cell surface is optimal for reducing interactions between nearby receptors and therefore maximizes ligand intake by the cell, where ligands were considered spatially homogeneous in the environment. Here, we consider the detection of spatial asymmetry of ligand concentrations around the cell due to the chemical gradient. By examining the effect of geometric layout of receptor aggregates on the cell surface, we showed that anisotropic receptor aggregate distribution generates a trade-off between the sensing limit of gradient steepness $p$ and that of the gradient direction $\phi$ (Fig.\ref{fig:fig3}). The isotropic layout represents an optimal configuration, in which the sensing accuracy is insensitive to cell orientation.

In summary, we study the effect of receptor cooperativity and adaptation on the chemical sensing limit by evaluating the Cramer-Rao lower bounds from the instantaneous global state of receptor activity. Our results showed that receptor cooperativity with receptor adaptation increases gradient sensing accuracy (by lowering the CRLB) for small signal changes across a wide dynamic range of background concentration. This result is also applicable to concentration sensing at a single aggregate location (Eq.~(\ref{eq:cm})). It remains largely unknown whether or how a chemotactic cell achieves its sensing limit. Receptor internalization~\cite{aquino2010increased} might improve chemical sensing by helping the cell membrane to function as an absorbing surface, an ideal device that operates at the fundamental Berg-Purcell limit. Maximum likelihood estimation (MLE) from ligand-receptor binding time series~\cite{endres2009maximum,*mora2010limits} or instantaneous receptor states~\cite{hu2010physical,hu2011geometry} was also suggested as a possible approach to the physical limit. The answer to how a cell mechanistically integrates information in time and space remains speculative and requires further experimental and theoretical investigations.

\section{Acknowledgement}
We thank Qiang Chang, Byron Goldstein, Libo Huang and Zhen Wang for many helpful discussions. The work was supported by National Science Foundation of China through grant 30870477.


\begin{thebibliography}{35}%
\makeatletter
\providecommand \@ifxundefined [1]{%
 \@ifx{#1\undefined}
}%
\providecommand \@ifnum [1]{%
 \ifnum #1\expandafter \@firstoftwo
 \else \expandafter \@secondoftwo
 \fi
}%
\providecommand \@ifx [1]{%
 \ifx #1\expandafter \@firstoftwo
 \else \expandafter \@secondoftwo
 \fi
}%
\providecommand \natexlab [1]{#1}%
\providecommand \enquote  [1]{``#1''}%
\providecommand \bibnamefont  [1]{#1}%
\providecommand \bibfnamefont [1]{#1}%
\providecommand \citenamefont [1]{#1}%
\providecommand \href@noop [0]{\@secondoftwo}%
\providecommand \href [0]{\begingroup \@sanitize@url \@href}%
\providecommand \@href[1]{\@@startlink{#1}\@@href}%
\providecommand \@@href[1]{\endgroup#1\@@endlink}%
\providecommand \@sanitize@url [0]{\catcode `\\12\catcode `\$12\catcode
  `\&12\catcode `\#12\catcode `\^12\catcode `\_12\catcode `\%12\relax}%
\providecommand \@@startlink[1]{}%
\providecommand \@@endlink[0]{}%
\providecommand \url  [0]{\begingroup\@sanitize@url \@url }%
\providecommand \@url [1]{\endgroup\@href {#1}{\urlprefix }}%
\providecommand \urlprefix  [0]{URL }%
\providecommand \Eprint [0]{\href }%
\providecommand \doibase [0]{http://dx.doi.org/}%
\providecommand \selectlanguage [0]{\@gobble}%
\providecommand \bibinfo  [0]{\@secondoftwo}%
\providecommand \bibfield  [0]{\@secondoftwo}%
\providecommand \translation [1]{[#1]}%
\providecommand \BibitemOpen [0]{}%
\providecommand \bibitemStop [0]{}%
\providecommand \bibitemNoStop [0]{.\EOS\space}%
\providecommand \EOS [0]{\spacefactor3000\relax}%
\providecommand \BibitemShut  [1]{\csname bibitem#1\endcsname}%
\let\auto@bib@innerbib\@empty
\bibitem [{\citenamefont {Berg}\ and\ \citenamefont
  {Purcell}(1977)}]{berg1977physics}%
  \BibitemOpen
  \bibfield  {author} {\bibinfo {author} {\bibfnamefont {H.~C.}\ \bibnamefont
  {Berg}}\ and\ \bibinfo {author} {\bibfnamefont {E.~M.}\ \bibnamefont
  {Purcell}},\ }\href@noop {} {\bibfield  {journal} {\bibinfo  {journal}
  {Biophys. J.}\ }\textbf {\bibinfo {volume} {20}},\ \bibinfo {pages} {193}
  (\bibinfo {year} {1977})}\BibitemShut {NoStop}%
\bibitem [{\citenamefont {Bialek}\ and\ \citenamefont
  {Setayeshgar}(2005)}]{bialek2005physical}%
  \BibitemOpen
  \bibfield  {author} {\bibinfo {author} {\bibfnamefont {W.}~\bibnamefont
  {Bialek}}\ and\ \bibinfo {author} {\bibfnamefont {S.}~\bibnamefont
  {Setayeshgar}},\ }\href@noop {} {\bibfield  {journal} {\bibinfo  {journal}
  {Proc. Natl. Acad. Sci. USA}\ }\textbf {\bibinfo {volume} {102}},\ \bibinfo
  {pages} {10040} (\bibinfo {year} {2005})}\BibitemShut {NoStop}%
\bibitem [{\citenamefont {Endres}\ and\ \citenamefont
  {Wingreen}(2008)}]{endres2008accuracy}%
  \BibitemOpen
  \bibfield  {author} {\bibinfo {author} {\bibfnamefont {R.~G.}\ \bibnamefont
  {Endres}}\ and\ \bibinfo {author} {\bibfnamefont {N.~S.}\ \bibnamefont
  {Wingreen}},\ }\href@noop {} {\bibfield  {journal} {\bibinfo  {journal}
  {Proc. Natl. Acad. Sci. USA}\ }\textbf {\bibinfo {volume} {105}},\ \bibinfo
  {pages} {15749} (\bibinfo {year} {2008})}\BibitemShut {NoStop}%
\bibitem [{\citenamefont {Endres}\ and\ \citenamefont
  {Wingreen}(2009{\natexlab{a}})}]{endres2009accuracy}%
  \BibitemOpen
  \bibfield  {author} {\bibinfo {author} {\bibfnamefont {R.~G.}\ \bibnamefont
  {Endres}}\ and\ \bibinfo {author} {\bibfnamefont {N.~S.}\ \bibnamefont
  {Wingreen}},\ }\href@noop {} {\bibfield  {journal} {\bibinfo  {journal}
  {Prog. Biophys. Mol. Biol.}\ }\textbf {\bibinfo {volume} {100}},\ \bibinfo
  {pages} {33} (\bibinfo {year} {2009}{\natexlab{a}})}\BibitemShut {NoStop}%
\bibitem [{\citenamefont {Bialek}\ and\ \citenamefont
  {Setayeshgar}(2008)}]{bialek2008cooperativity}%
  \BibitemOpen
  \bibfield  {author} {\bibinfo {author} {\bibfnamefont {W.}~\bibnamefont
  {Bialek}}\ and\ \bibinfo {author} {\bibfnamefont {S.}~\bibnamefont
  {Setayeshgar}},\ }\href@noop {} {\bibfield  {journal} {\bibinfo  {journal}
  {Phys. Rev. Lett.}\ }\textbf {\bibinfo {volume} {100}},\ \bibinfo {pages}
  {258101} (\bibinfo {year} {2008})}\BibitemShut {NoStop}%
\bibitem [{\citenamefont {Hu}\ \emph {et~al.}(2010)\citenamefont {Hu},
  \citenamefont {Chen}, \citenamefont {Rappel},\ and\ \citenamefont
  {Levine}}]{hu2010physical}%
  \BibitemOpen
  \bibfield  {author} {\bibinfo {author} {\bibfnamefont {B.}~\bibnamefont
  {Hu}}, \bibinfo {author} {\bibfnamefont {W.}~\bibnamefont {Chen}}, \bibinfo
  {author} {\bibfnamefont {W.~J.}\ \bibnamefont {Rappel}}, \ and\ \bibinfo
  {author} {\bibfnamefont {H.}~\bibnamefont {Levine}},\ }\href@noop {}
  {\bibfield  {journal} {\bibinfo  {journal} {Phys. Rev. Lett.}\ }\textbf
  {\bibinfo {volume} {105}},\ \bibinfo {pages} {48104} (\bibinfo {year}
  {2010})}\BibitemShut {NoStop}%
\bibitem [{\citenamefont {Aquino}\ \emph {et~al.}(2011)\citenamefont {Aquino},
  \citenamefont {Clausznitzer}, \citenamefont {Tollis},\ and\ \citenamefont
  {Endres}}]{aquino2011optimal}%
  \BibitemOpen
  \bibfield  {author} {\bibinfo {author} {\bibfnamefont {G.}~\bibnamefont
  {Aquino}}, \bibinfo {author} {\bibfnamefont {D.}~\bibnamefont
  {Clausznitzer}}, \bibinfo {author} {\bibfnamefont {S.}~\bibnamefont
  {Tollis}}, \ and\ \bibinfo {author} {\bibfnamefont {R.~G.}\ \bibnamefont
  {Endres}},\ }\href@noop {} {\bibfield  {journal} {\bibinfo  {journal} {Phys.
  Rev. E}\ }\textbf {\bibinfo {volume} {83}},\ \bibinfo {pages} {021914}
  (\bibinfo {year} {2011})}\BibitemShut {NoStop}%
\bibitem [{\citenamefont {Skoge}\ \emph {et~al.}(2011)\citenamefont {Skoge},
  \citenamefont {Meir},\ and\ \citenamefont {Wingreen}}]{skoge2011dynamics}%
  \BibitemOpen
  \bibfield  {author} {\bibinfo {author} {\bibfnamefont {M.}~\bibnamefont
  {Skoge}}, \bibinfo {author} {\bibfnamefont {Y.}~\bibnamefont {Meir}}, \ and\
  \bibinfo {author} {\bibfnamefont {N.~S.}\ \bibnamefont {Wingreen}},\
  }\href@noop {} {\bibfield  {journal} {\bibinfo  {journal} {Phys. Rev. Lett.}\
  }\textbf {\bibinfo {volume} {105}},\ \bibinfo {pages} {178101} (\bibinfo
  {year} {2011})}\BibitemShut {NoStop}%
\bibitem [{\citenamefont {Monod}\ \emph {et~al.}(1965)\citenamefont {Monod},
  \citenamefont {Wyman},\ and\ \citenamefont {Changeux}}]{monod1965nature}%
  \BibitemOpen
  \bibfield  {author} {\bibinfo {author} {\bibfnamefont {J.}~\bibnamefont
  {Monod}}, \bibinfo {author} {\bibfnamefont {J.}~\bibnamefont {Wyman}}, \ and\
  \bibinfo {author} {\bibfnamefont {J.~P.}\ \bibnamefont {Changeux}},\
  }\href@noop {} {\bibfield  {journal} {\bibinfo  {journal} {J. Mol. Biol.}\
  }\textbf {\bibinfo {volume} {12}},\ \bibinfo {pages} {88} (\bibinfo {year}
  {1965})}\BibitemShut {NoStop}%
\bibitem [{\citenamefont {Changeux}(2012)}]{changeux2012}%
  \BibitemOpen
  \bibfield  {author} {\bibinfo {author} {\bibfnamefont {J.~P.}\ \bibnamefont
  {Changeux}},\ }\href@noop {} {\bibfield  {journal} {\bibinfo  {journal}
  {Annu. Rev. Biophys.}\ }\textbf {\bibinfo {volume} {41}} (\bibinfo {year}
  {2012})}\BibitemShut {NoStop}%
\bibitem [{\citenamefont {Sourjik}\ and\ \citenamefont
  {Berg}(2004)}]{sourjik2004functional}%
  \BibitemOpen
  \bibfield  {author} {\bibinfo {author} {\bibfnamefont {V.}~\bibnamefont
  {Sourjik}}\ and\ \bibinfo {author} {\bibfnamefont {H.~C.}\ \bibnamefont
  {Berg}},\ }\href@noop {} {\bibfield  {journal} {\bibinfo  {journal} {Nature}\
  }\textbf {\bibinfo {volume} {428}},\ \bibinfo {pages} {437} (\bibinfo {year}
  {2004})}\BibitemShut {NoStop}%
\bibitem [{\citenamefont {Mello}\ and\ \citenamefont
  {Tu}(2005)}]{mello2005allosteric}%
  \BibitemOpen
  \bibfield  {author} {\bibinfo {author} {\bibfnamefont {B.~A.}\ \bibnamefont
  {Mello}}\ and\ \bibinfo {author} {\bibfnamefont {Y.}~\bibnamefont {Tu}},\
  }\href@noop {} {\bibfield  {journal} {\bibinfo  {journal} {Proc. Natl. Acad.
  Sci. USA}\ }\textbf {\bibinfo {volume} {102}},\ \bibinfo {pages} {17354}
  (\bibinfo {year} {2005})}\BibitemShut {NoStop}%
\bibitem [{\citenamefont {Keymer}\ \emph {et~al.}(2006)\citenamefont {Keymer},
  \citenamefont {Endres}, \citenamefont {Skoge}, \citenamefont {Meir},\ and\
  \citenamefont {Wingreen}}]{keymer2006chemosensing}%
  \BibitemOpen
  \bibfield  {author} {\bibinfo {author} {\bibfnamefont {J.~E.}\ \bibnamefont
  {Keymer}}, \bibinfo {author} {\bibfnamefont {R.~G.}\ \bibnamefont {Endres}},
  \bibinfo {author} {\bibfnamefont {M.}~\bibnamefont {Skoge}}, \bibinfo
  {author} {\bibfnamefont {Y.}~\bibnamefont {Meir}}, \ and\ \bibinfo {author}
  {\bibfnamefont {N.~S.}\ \bibnamefont {Wingreen}},\ }\href@noop {} {\bibfield
  {journal} {\bibinfo  {journal} {Proc. Natl. Acad. Sci. USA}\ }\textbf
  {\bibinfo {volume} {103}},\ \bibinfo {pages} {1786} (\bibinfo {year}
  {2006})}\BibitemShut {NoStop}%
\bibitem [{\citenamefont {Skoge}\ \emph {et~al.}(2006)\citenamefont {Skoge},
  \citenamefont {Endres},\ and\ \citenamefont {Wingreen}}]{skoge2006receptor}%
  \BibitemOpen
  \bibfield  {author} {\bibinfo {author} {\bibfnamefont {M.~L.}\ \bibnamefont
  {Skoge}}, \bibinfo {author} {\bibfnamefont {R.~G.}\ \bibnamefont {Endres}}, \
  and\ \bibinfo {author} {\bibfnamefont {N.~S.}\ \bibnamefont {Wingreen}},\
  }\href@noop {} {\bibfield  {journal} {\bibinfo  {journal} {Biophys. J.}\
  }\textbf {\bibinfo {volume} {90}},\ \bibinfo {pages} {4317} (\bibinfo {year}
  {2006})}\BibitemShut {NoStop}%
\bibitem [{\citenamefont {Kay}(1993)}]{kay1993fundamentals}%
  \BibitemOpen
  \bibfield  {author} {\bibinfo {author} {\bibfnamefont {S.~M.}\ \bibnamefont
  {Kay}},\ }\href@noop {} {\emph {\bibinfo {title} {Fundamentals of Statistical
  Signal Processing, Vol. I: Estimation Theory}}}\ (\bibinfo  {publisher}
  {Prentice Hall},\ \bibinfo {year} {1993})\BibitemShut {NoStop}%
\bibitem [{Note1()}]{Note1}%
  \BibitemOpen
  \bibinfo {note} {For an equidistant layout of receptor aggregates ($M\ge 3$),
  $\DOTSB \sum@ \slimits@ _{m=1}^M\protect \qopname \relax o{cos}^2(\theta
  _m-\phi )=\DOTSB \sum@ \slimits@ _{m=1}^M\protect \qopname \relax
  o{sin}^2(\theta _m-\phi )=M/2$ and $\DOTSB \sum@ \slimits@ _{m=1}^M\protect
  \qopname \relax o{sin}2(\theta _m-\phi )=0$.}\BibitemShut {Stop}%
\bibitem [{Note2()}]{Note2}%
  \BibitemOpen
  \bibinfo {note} {In this scenario, the factor related cooperativity is $\xi
  _n=n^2P_0(1-P_0)$. $\partial \xi _n/\partial n>0$ implies that
  $(1-2P_0)\protect \qopname \relax o{ln}(1/P_0-1)<2$, or, $P_-<P_0<P_+$, where
  $P_- \approx 0.083$ and $P_+\approx 0.917$.}\BibitemShut {Stop}%
\bibitem [{\citenamefont {Berg}\ and\ \citenamefont
  {Brown}(1972)}]{berg1972chemotaxis}%
  \BibitemOpen
  \bibfield  {author} {\bibinfo {author} {\bibfnamefont {H.~C.}\ \bibnamefont
  {Berg}}\ and\ \bibinfo {author} {\bibfnamefont {D.~A.}\ \bibnamefont
  {Brown}},\ }\href@noop {} {\bibfield  {journal} {\bibinfo  {journal}
  {Nature}\ }\textbf {\bibinfo {volume} {239}},\ \bibinfo {pages} {500}
  (\bibinfo {year} {1972})}\BibitemShut {NoStop}%
\bibitem [{\citenamefont {Mello}\ and\ \citenamefont
  {Tu}(2007)}]{mello2007effects}%
  \BibitemOpen
  \bibfield  {author} {\bibinfo {author} {\bibfnamefont {B.~A.}\ \bibnamefont
  {Mello}}\ and\ \bibinfo {author} {\bibfnamefont {Y.}~\bibnamefont {Tu}},\
  }\href@noop {} {\bibfield  {journal} {\bibinfo  {journal} {Biophys. J.}\
  }\textbf {\bibinfo {volume} {92}},\ \bibinfo {pages} {2329} (\bibinfo {year}
  {2007})}\BibitemShut {NoStop}%
\bibitem [{\citenamefont {Clausznitzer}\ \emph {et~al.}(2010)\citenamefont
  {Clausznitzer}, \citenamefont {Oleksiuk}, \citenamefont {L{\o}vdok},
  \citenamefont {Sourjik},\ and\ \citenamefont
  {Endres}}]{clausznitzer2010chemotactic}%
  \BibitemOpen
  \bibfield  {author} {\bibinfo {author} {\bibfnamefont {D.}~\bibnamefont
  {Clausznitzer}}, \bibinfo {author} {\bibfnamefont {O.}~\bibnamefont
  {Oleksiuk}}, \bibinfo {author} {\bibfnamefont {L.}~\bibnamefont {L{\o}vdok}},
  \bibinfo {author} {\bibfnamefont {V.}~\bibnamefont {Sourjik}}, \ and\
  \bibinfo {author} {\bibfnamefont {R.}~\bibnamefont {Endres}},\ }\href@noop {}
  {\bibfield  {journal} {\bibinfo  {journal} {PLoS Comp. Biol.}\ }\textbf
  {\bibinfo {volume} {6}},\ \bibinfo {pages} {e1000784} (\bibinfo {year}
  {2010})}\BibitemShut {NoStop}%
\bibitem [{\citenamefont {Hu}\ \emph {et~al.}(2011)\citenamefont {Hu},
  \citenamefont {Chen}, \citenamefont {Rappel},\ and\ \citenamefont
  {Levine}}]{hu2011geometry}%
  \BibitemOpen
  \bibfield  {author} {\bibinfo {author} {\bibfnamefont {B.}~\bibnamefont
  {Hu}}, \bibinfo {author} {\bibfnamefont {W.}~\bibnamefont {Chen}}, \bibinfo
  {author} {\bibfnamefont {W.~J.}\ \bibnamefont {Rappel}}, \ and\ \bibinfo
  {author} {\bibfnamefont {H.}~\bibnamefont {Levine}},\ }\href@noop {}
  {\bibfield  {journal} {\bibinfo  {journal} {Phys. Rev. E}\ }\textbf {\bibinfo
  {volume} {83}},\ \bibinfo {pages} {021917} (\bibinfo {year}
  {2011})}\BibitemShut {NoStop}%
\bibitem [{\citenamefont {Wang}\ \emph {et~al.}(2007)\citenamefont {Wang},
  \citenamefont {Rappel}, \citenamefont {Kerr},\ and\ \citenamefont
  {Levine}}]{wang2007quantifying}%
  \BibitemOpen
  \bibfield  {author} {\bibinfo {author} {\bibfnamefont {K.}~\bibnamefont
  {Wang}}, \bibinfo {author} {\bibfnamefont {W.~J.}\ \bibnamefont {Rappel}},
  \bibinfo {author} {\bibfnamefont {R.}~\bibnamefont {Kerr}}, \ and\ \bibinfo
  {author} {\bibfnamefont {H.}~\bibnamefont {Levine}},\ }\href@noop {}
  {\bibfield  {journal} {\bibinfo  {journal} {Phys. Rev. E}\ }\textbf {\bibinfo
  {volume} {75}},\ \bibinfo {pages} {061905} (\bibinfo {year}
  {2007})}\BibitemShut {NoStop}%
\bibitem [{\citenamefont {Glauber}(1963)}]{glauber1963time}%
  \BibitemOpen
  \bibfield  {author} {\bibinfo {author} {\bibfnamefont {R.~J.}\ \bibnamefont
  {Glauber}},\ }\href@noop {} {\bibfield  {journal} {\bibinfo  {journal} {J.
  Math. phys.}\ }\textbf {\bibinfo {volume} {4}},\ \bibinfo {pages} {294}
  (\bibinfo {year} {1963})}\BibitemShut {NoStop}%
\bibitem [{Note3()}]{Note3}%
  \BibitemOpen
  \bibinfo {note} {The effective rate constants under ligand rebinding are
  given as: $\protect \mathaccentV {tilde}07E{k}_f=k_f/(1+[R]k_f/k_+)$ and
  $\protect \mathaccentV {tilde}07E{k}_r=k_r/(1+[R]k_f/k_+)$, where $k_+=4Da$
  is the diffusion limited on rate to a receptor aggregate of size $a$.
  $[R]k_f/k_+$ is the rebinding/escape ratio that characterizes the degree of
  influence by ligand rebinding.}\BibitemShut {Stop}%
\bibitem [{\citenamefont {Shoup}\ and\ \citenamefont
  {Szabo}(1982)}]{shoup1982role}%
  \BibitemOpen
  \bibfield  {author} {\bibinfo {author} {\bibfnamefont {D.}~\bibnamefont
  {Shoup}}\ and\ \bibinfo {author} {\bibfnamefont {A.}~\bibnamefont {Szabo}},\
  }\href@noop {} {\bibfield  {journal} {\bibinfo  {journal} {Biophys. J.}\
  }\textbf {\bibinfo {volume} {40}},\ \bibinfo {pages} {33} (\bibinfo {year}
  {1982})}\BibitemShut {NoStop}%
\bibitem [{\citenamefont {Goldstein}\ and\ \citenamefont
  {Dembo}(1995)}]{goldstein1995approximating}%
  \BibitemOpen
  \bibfield  {author} {\bibinfo {author} {\bibfnamefont {B.}~\bibnamefont
  {Goldstein}}\ and\ \bibinfo {author} {\bibfnamefont {M.}~\bibnamefont
  {Dembo}},\ }\href@noop {} {\bibfield  {journal} {\bibinfo  {journal}
  {Biophys. J.}\ }\textbf {\bibinfo {volume} {68}},\ \bibinfo {pages} {1222}
  (\bibinfo {year} {1995})}\BibitemShut {NoStop}%
\bibitem [{\citenamefont {Andrews}(2005)}]{andrews2005serial}%
  \BibitemOpen
  \bibfield  {author} {\bibinfo {author} {\bibfnamefont {S.~S.}\ \bibnamefont
  {Andrews}},\ }\href@noop {} {\bibfield  {journal} {\bibinfo  {journal} {Phys.
  Biol.}\ }\textbf {\bibinfo {volume} {2}},\ \bibinfo {pages} {111} (\bibinfo
  {year} {2005})}\BibitemShut {NoStop}%
\bibitem [{\citenamefont {Briegel}\ \emph {et~al.}(2009)\citenamefont
  {Briegel}, \citenamefont {Ortega}, \citenamefont {Tocheva}, \citenamefont
  {Wuichet} \emph {et~al.}}]{briegel2009universal}%
  \BibitemOpen
  \bibfield  {author} {\bibinfo {author} {\bibfnamefont {A.}~\bibnamefont
  {Briegel}}, \bibinfo {author} {\bibfnamefont {D.}~\bibnamefont {Ortega}},
  \bibinfo {author} {\bibfnamefont {E.~I.}\ \bibnamefont {Tocheva}}, \bibinfo
  {author} {\bibfnamefont {K.}~\bibnamefont {Wuichet}},  \emph {et~al.},\
  }\href@noop {} {\bibfield  {journal} {\bibinfo  {journal} {Proc. Natl. Acad.
  Sci. USA}\ }\textbf {\bibinfo {volume} {106}},\ \bibinfo {pages} {17181}
  (\bibinfo {year} {2009})}\BibitemShut {NoStop}%
\bibitem [{\citenamefont {Briegel}\ \emph {et~al.}(2011)\citenamefont
  {Briegel}, \citenamefont {Beeby}, \citenamefont {Thanbichler}, \citenamefont
  {Jensen} \emph {et~al.}}]{briegel2011activated}%
  \BibitemOpen
  \bibfield  {author} {\bibinfo {author} {\bibfnamefont {A.}~\bibnamefont
  {Briegel}}, \bibinfo {author} {\bibfnamefont {M.}~\bibnamefont {Beeby}},
  \bibinfo {author} {\bibfnamefont {M.}~\bibnamefont {Thanbichler}}, \bibinfo
  {author} {\bibfnamefont {G.~J.}\ \bibnamefont {Jensen}},  \emph {et~al.},\
  }\href@noop {} {\bibfield  {journal} {\bibinfo  {journal} {Mol. Microbiol.}\
  }\textbf {\bibinfo {volume} {82}},\ \bibinfo {pages} {748} (\bibinfo {year}
  {2011})}\BibitemShut {NoStop}%
\bibitem [{\citenamefont {Briegel}\ \emph {et~al.}(2012)\citenamefont
  {Briegel}, \citenamefont {Li}, \citenamefont {Bilwes}, \citenamefont {Hughes}
  \emph {et~al.}}]{Briegel21022012}%
  \BibitemOpen
  \bibfield  {author} {\bibinfo {author} {\bibfnamefont {A.}~\bibnamefont
  {Briegel}}, \bibinfo {author} {\bibfnamefont {X.}~\bibnamefont {Li}},
  \bibinfo {author} {\bibfnamefont {A.~M.}\ \bibnamefont {Bilwes}}, \bibinfo
  {author} {\bibfnamefont {K.~T.}\ \bibnamefont {Hughes}},  \emph {et~al.},\
  }\href@noop {} {\bibfield  {journal} {\bibinfo  {journal} {Proc. Natl. Acad.
  Sci. USA}\ }\textbf {\bibinfo {volume} {109}},\ \bibinfo {pages} {3766}
  (\bibinfo {year} {2012})}\BibitemShut {NoStop}%
\bibitem [{\citenamefont {Liu}\ \emph {et~al.}(2012)\citenamefont {Liu},
  \citenamefont {Hu}, \citenamefont {Morado}, \citenamefont {Jani},
  \citenamefont {Manson},\ and\ \citenamefont {Margolin}}]{liu2012molecular}%
  \BibitemOpen
  \bibfield  {author} {\bibinfo {author} {\bibfnamefont {J.}~\bibnamefont
  {Liu}}, \bibinfo {author} {\bibfnamefont {B.}~\bibnamefont {Hu}}, \bibinfo
  {author} {\bibfnamefont {D.~R.}\ \bibnamefont {Morado}}, \bibinfo {author}
  {\bibfnamefont {S.}~\bibnamefont {Jani}}, \bibinfo {author} {\bibfnamefont
  {M.~D.}\ \bibnamefont {Manson}}, \ and\ \bibinfo {author} {\bibfnamefont
  {W.}~\bibnamefont {Margolin}},\ }\href@noop {} {\bibfield  {journal}
  {\bibinfo  {journal} {Proc. Natl. Acad. Sci. USA}\ }\textbf {\bibinfo
  {volume} {109}},\ \bibinfo {pages} {E1481} (\bibinfo {year}
  {2012})}\BibitemShut {NoStop}%
\bibitem [{\citenamefont {Wadhams}\ and\ \citenamefont
  {Armitage}(2004)}]{wadhams2004making}%
  \BibitemOpen
  \bibfield  {author} {\bibinfo {author} {\bibfnamefont {G.~H.}\ \bibnamefont
  {Wadhams}}\ and\ \bibinfo {author} {\bibfnamefont {J.~P.}\ \bibnamefont
  {Armitage}},\ }\href@noop {} {\bibfield  {journal} {\bibinfo  {journal} {Nat.
  Rev. Mol. Cell Biol.}\ }\textbf {\bibinfo {volume} {5}},\ \bibinfo {pages}
  {1024} (\bibinfo {year} {2004})}\BibitemShut {NoStop}%
\bibitem [{\citenamefont {Aquino}\ and\ \citenamefont
  {Endres}(2010)}]{aquino2010increased}%
  \BibitemOpen
  \bibfield  {author} {\bibinfo {author} {\bibfnamefont {G.}~\bibnamefont
  {Aquino}}\ and\ \bibinfo {author} {\bibfnamefont {R.~G.}\ \bibnamefont
  {Endres}},\ }\href@noop {} {\bibfield  {journal} {\bibinfo  {journal} {Phys.
  Rev. E}\ }\textbf {\bibinfo {volume} {81}},\ \bibinfo {pages} {21909}
  (\bibinfo {year} {2010})}\BibitemShut {NoStop}%
\bibitem [{\citenamefont {Endres}\ and\ \citenamefont
  {Wingreen}(2009{\natexlab{b}})}]{endres2009maximum}%
  \BibitemOpen
  \bibfield  {author} {\bibinfo {author} {\bibfnamefont {R.~G.}\ \bibnamefont
  {Endres}}\ and\ \bibinfo {author} {\bibfnamefont {N.~S.}\ \bibnamefont
  {Wingreen}},\ }\href@noop {} {\bibfield  {journal} {\bibinfo  {journal}
  {Phys. Rev. Lett.}\ }\textbf {\bibinfo {volume} {103}},\ \bibinfo {pages}
  {158101} (\bibinfo {year} {2009}{\natexlab{b}})}\BibitemShut {NoStop}%
\bibitem [{\citenamefont {Mora}\ and\ \citenamefont
  {Wingreen}(2010)}]{mora2010limits}%
  \BibitemOpen
  \bibfield  {author} {\bibinfo {author} {\bibfnamefont {T.}~\bibnamefont
  {Mora}}\ and\ \bibinfo {author} {\bibfnamefont {N.~S.}\ \bibnamefont
  {Wingreen}},\ }\href@noop {} {\bibfield  {journal} {\bibinfo  {journal}
  {Phys. Rev. Lett.}\ }\textbf {\bibinfo {volume} {104}},\ \bibinfo {pages}
  {248101} (\bibinfo {year} {2010})}\BibitemShut {NoStop}%
\end{thebibliography}
%

\end{document}